\documentclass[10pt, conference, letterpaper]{IEEEtran}        
\usepackage{graphicx}
\usepackage{mathptmx}      
\usepackage{subcaption} 
\usepackage{caption}
\usepackage{verbatim}
\usepackage{listings}
\usepackage{url}
\usepackage{array}
\usepackage{tcolorbox}
\usepackage{tikz}
\usepackage{mathrsfs}
\usepackage{bbm}
\usetikzlibrary{fit,arrows,calc,spy,decorations.pathreplacing,shapes, patterns}
\usepackage{stmaryrd} 
\usepackage{color}
\usepackage{amsmath} 
\usepackage[utf8]{inputenc}
\usepackage{amssymb} 
\usepackage{epstopdf}
\usepackage{keyval}
\usepackage{ifthen}
\usepackage{pgf-pie}

\makeatother
\makeatletter
\newcommand{\tikzcuboid@shiftx}{0}
\newcommand{\tikzcuboid@shifty}{0}
\newcommand{\tikzcuboid@dimx}{3}
\newcommand{\tikzcuboid@dimy}{3}
\newcommand{\tikzcuboid@dimz}{3}
\newcommand{\tikzcuboid@scale}{1}
\newcommand{\tikzcuboid@densityx}{1}
\newcommand{\tikzcuboid@densityy}{1}
\newcommand{\tikzcuboid@densityz}{1}
\newcommand{\tikzcuboid@rotation}{0}
\newcommand{\tikzcuboid@anglex}{0}
\newcommand{\tikzcuboid@angley}{90}
\newcommand{\tikzcuboid@anglez}{225}
\newcommand{\tikzcuboid@scalex}{1}
\newcommand{\tikzcuboid@scaley}{1}
\newcommand{\tikzcuboid@scalez}{sqrt(0.5)}
\newcommand{\tikzcuboid@linefront}{black}
\newcommand{\tikzcuboid@linetop}{black}
\newcommand{\tikzcuboid@lineright}{black}
\newcommand{\tikzcuboid@fillfront}{white}
\newcommand{\tikzcuboid@filltop}{white}
\newcommand{\tikzcuboid@fillright}{white}
\newcommand{\tikzcuboid@shaded}{N}
\newcommand{\tikzcuboid@shadecolor}{black}
\newcommand{\tikzcuboid@shadeperc}{25}
\newcommand{\tikzcuboid@emphedge}{N}
\newcommand{\tikzcuboid@emphstyle}{thick}

\define@key{tikzcuboid}{shiftx}[\tikzcuboid@shiftx]{\renewcommand{\tikzcuboid@shiftx}{#1}}
\define@key{tikzcuboid}{shifty}[\tikzcuboid@shifty]{\renewcommand{\tikzcuboid@shifty}{#1}}
\define@key{tikzcuboid}{dimx}[\tikzcuboid@dimx]{\renewcommand{\tikzcuboid@dimx}{#1}}
\define@key{tikzcuboid}{dimy}[\tikzcuboid@dimy]{\renewcommand{\tikzcuboid@dimy}{#1}}
\define@key{tikzcuboid}{dimz}[\tikzcuboid@dimz]{\renewcommand{\tikzcuboid@dimz}{#1}}
\define@key{tikzcuboid}{scale}[\tikzcuboid@scale]{\renewcommand{\tikzcuboid@scale}{#1}}
\define@key{tikzcuboid}{densityx}[\tikzcuboid@densityx]{\renewcommand{\tikzcuboid@densityx}{#1}}
\define@key{tikzcuboid}{densityy}[\tikzcuboid@densityy]{\renewcommand{\tikzcuboid@densityy}{#1}}
\define@key{tikzcuboid}{densityz}[\tikzcuboid@densityz]{\renewcommand{\tikzcuboid@densityz}{#1}}
\define@key{tikzcuboid}{rotation}[\tikzcuboid@rotation]{\renewcommand{\tikzcuboid@rotation}{#1}}
\define@key{tikzcuboid}{anglex}[\tikzcuboid@anglex]{\renewcommand{\tikzcuboid@anglex}{#1}}
\define@key{tikzcuboid}{angley}[\tikzcuboid@angley]{\renewcommand{\tikzcuboid@angley}{#1}}
\define@key{tikzcuboid}{anglez}[\tikzcuboid@anglez]{\renewcommand{\tikzcuboid@anglez}{#1}}
\define@key{tikzcuboid}{scalex}[\tikzcuboid@scalex]{\renewcommand{\tikzcuboid@scalex}{#1}}
\define@key{tikzcuboid}{scaley}[\tikzcuboid@scaley]{\renewcommand{\tikzcuboid@scaley}{#1}}
\define@key{tikzcuboid}{scalez}[\tikzcuboid@scalez]{\renewcommand{\tikzcuboid@scalez}{#1}}
\define@key{tikzcuboid}{linefront}[\tikzcuboid@linefront]{\renewcommand{\tikzcuboid@linefront}{#1}}
\define@key{tikzcuboid}{linetop}[\tikzcuboid@linetop]{\renewcommand{\tikzcuboid@linetop}{#1}}
\define@key{tikzcuboid}{lineright}[\tikzcuboid@lineright]{\renewcommand{\tikzcuboid@lineright}{#1}}
\define@key{tikzcuboid}{fillfront}[\tikzcuboid@fillfront]{\renewcommand{\tikzcuboid@fillfront}{#1}}
\define@key{tikzcuboid}{filltop}[\tikzcuboid@filltop]{\renewcommand{\tikzcuboid@filltop}{#1}}
\define@key{tikzcuboid}{fillright}[\tikzcuboid@fillright]{\renewcommand{\tikzcuboid@fillright}{#1}}
\define@key{tikzcuboid}{shaded}[\tikzcuboid@shaded]{\renewcommand{\tikzcuboid@shaded}{#1}}
\define@key{tikzcuboid}{shadecolor}[\tikzcuboid@shadecolor]{\renewcommand{\tikzcuboid@shadecolor}{#1}}
\define@key{tikzcuboid}{shadeperc}[\tikzcuboid@shadeperc]{\renewcommand{\tikzcuboid@shadeperc}{#1}}
\define@key{tikzcuboid}{emphedge}[\tikzcuboid@emphedge]{\renewcommand{\tikzcuboid@emphedge}{#1}}
\define@key{tikzcuboid}{emphstyle}[\tikzcuboid@emphstyle]{\renewcommand{\tikzcuboid@emphstyle}{#1}}

\newcommand{\tikzcuboid}[1]{
    \setkeys{tikzcuboid}{#1} 
    \pgfmathsetmacro{\vectorxx}{\tikzcuboid@scalex*cos(\tikzcuboid@anglex)}
    \pgfmathsetmacro{\vectorxy}{\tikzcuboid@scalex*sin(\tikzcuboid@anglex)}
    \pgfmathsetmacro{\vectoryx}{\tikzcuboid@scaley*cos(\tikzcuboid@angley)}
    \pgfmathsetmacro{\vectoryy}{\tikzcuboid@scaley*sin(\tikzcuboid@angley)}
    \pgfmathsetmacro{\vectorzx}{\tikzcuboid@scalez*cos(\tikzcuboid@anglez)}
    \pgfmathsetmacro{\vectorzy}{\tikzcuboid@scalez*sin(\tikzcuboid@anglez)}
    \begin{scope}[xshift=\tikzcuboid@shiftx, yshift=\tikzcuboid@shifty, scale=\tikzcuboid@scale, rotate=\tikzcuboid@rotation, x={(\vectorxx,\vectorxy)}, y={(\vectoryx,\vectoryy)}, z={(\vectorzx,\vectorzy)}]
    \pgfmathsetmacro{\steppingx}{1/\tikzcuboid@densityx}
    \pgfmathsetmacro{\steppingy}{1/\tikzcuboid@densityy}
    \pgfmathsetmacro{\steppingz}{1/\tikzcuboid@densityz}
    \newcommand{\dimx}{\tikzcuboid@dimx}
    \newcommand{\dimy}{\tikzcuboid@dimy}
    \newcommand{\dimz}{\tikzcuboid@dimz}
    \pgfmathsetmacro{\secondx}{2*\steppingx}
    \pgfmathsetmacro{\secondy}{2*\steppingy}
    \pgfmathsetmacro{\secondz}{2*\steppingz}

             \ifthenelse{\equal{\dimx}{1}}
    {\foreach \x in {\steppingx,...,\dimx}}
    {\foreach \x in {\steppingx,\secondx,...,\dimx}}
  {     \ifthenelse{\equal{\dimy}{1}}
    {\foreach \y in {\steppingy,...,\dimy}}
    {\foreach \y in {\steppingy,\secondy,...,\dimy}}
    {   \pgfmathsetmacro{\lowx}{(\x-\steppingx)}
      \pgfmathsetmacro{\lowy}{(\y-\steppingy)}
      \filldraw[fill=\tikzcuboid@fillfront,draw=\tikzcuboid@linefront] (\lowx,\lowy,\dimz) -- (\lowx,\y,\dimz) -- (\x,\y,\dimz) -- (\x,\lowy,\dimz) -- cycle;
    }
    }
    \ifthenelse{\equal{\dimx}{1}}
    {\foreach \x in {\steppingx,...,\dimx}}
    {\foreach \x in {\steppingx,\secondx,...,\dimx}}
  { \ifthenelse{\equal{\dimz}{1}}
    {\foreach \z in {\steppingz,...,\dimz}}
    {\foreach \z in {\steppingz,\secondz,...,\dimz}}
    {   \pgfmathsetmacro{\lowx}{(\x-\steppingx)}
      \pgfmathsetmacro{\lowz}{(\z-\steppingz)}
      \filldraw[fill=\tikzcuboid@filltop,draw=\tikzcuboid@linetop] (\lowx,\dimy,\lowz) -- (\lowx,\dimy,\z) -- (\x,\dimy,\z) -- (\x,\dimy,\lowz) -- cycle;
        }
    }
    \ifthenelse{\equal{\dimy}{1}}
    {\foreach \y in {\steppingy,...,\dimy}}
    {\foreach \y in {\steppingy,\secondy,...,\dimy}}
  { \ifthenelse{\equal{\dimz}{1}}
    {\foreach \z in {\steppingz,...,\dimz}}
    {\foreach \z in {\steppingz,\secondz,...,\dimz}}
    {   \pgfmathsetmacro{\lowy}{(\y-\steppingy)}
      \pgfmathsetmacro{\lowz}{(\z-\steppingz)}
     \filldraw[fill=\tikzcuboid@fillright,draw=\tikzcuboid@lineright] (\dimx,\lowy,\lowz) -- (\dimx,\lowy,\z) -- (\dimx,\y,\z) -- (\dimx,\y,\lowz) -- cycle;
    }
  }

    \end{scope}
}

\tikzset{pics/.cd,
  pic a/.style={code={
      \node [fill=red!75, shape=circle] {A};
  }}
}

\newcommand{\centered}[1]{\begin{tabular}{l} #1 \end{tabular}}

\makeatother

\begin{document}

\title{Multidimensional Outlier Detection in Temporal Interaction Networks: An Application to Political Communication on Twitter
\thanks{This article is a substantially extended and revised version of the authors' COMPLENET 2019 paper \cite{wilmet2019multidimensional}, with an updated research, literature review and methodology, along with new data analysis.
}
}


\author{
\IEEEauthorblockN{Audrey Wilmet\IEEEauthorrefmark{1}, Robin Lamarche-Perrin\IEEEauthorrefmark{2}}
\IEEEauthorblockA{\IEEEauthorrefmark{1}Sorbonne Universit{\'e}, CNRS, Laboratoire d'Informatique de Paris 6, LIP6, F-75005 Paris, France}
\IEEEauthorblockA{\IEEEauthorrefmark{2}Institut des Systèmes Complexes de Paris Île-de-France, ISC-PIF, UPS 3611, Paris, France\\
Email: firstname.lastname@lip6.fr }
}




\date{}

\maketitle

\begin{abstract}
In social network Twitter, users can interact with each other and spread information via retweets. These millions of interactions may result in media events whose influence goes beyond Twitter framework. In this paper, we thoroughly explore interactions to provide a better understanding of the emergence of certain trends. First, we consider an interaction on Twitter to be a triplet $(s,a,t)$ meaning that user $s$, called the spreader, has retweeted a tweet of user $a$, called the author, at time $t$. We model this set of interactions as a data cube with three dimensions: spreaders, authors and time. Then, we provide a method which builds different contexts, where a context is a set of features characterizing the circumstances of an event. Finally, these contexts allow us to find relevant unexpected behaviors, according to several dimensions and various perspectives: a user during a given hour which is abnormal compared to its usual behavior, a relationship between two users which is abnormal compared to all other relationships, \textit{etc.} We apply our method to a set of retweets related to the 2017 French presidential election and show that one can build interesting insights regarding political organization on Twitter.
\end{abstract}

\section{Introduction}
The use of social networks has exploded over the past fifteen years. The micro-blogging service Twitter is currently the most popular and fastest-growing one of them. Within this social network, users can post information via tweets as well as spread information by retweeting tweets of other users. This leads to a dissemination of information from a variety of perspectives, thus affecting users ideas and opinions.\\
As discussed in the works of Murthy et al. \cite{murthy2013twitter} and Weller et al. \cite{weller2014twitter}, for some of the most active users, Twitter even constitute the primary medium by which they get informed. These users only represent a negligible fraction of the population. Nevertheless, hot topics emerging on Twitter's data stream are relayed by traditional media and therefore reach a much broader audience. If such trends often naturally arise from discussions or are consequences of the reaction of all users to real-world events, they may also be originated by the intensive activity of a small group and mislead other users on the significance of certain topics. \\
The volume of user-generated data is considerable: over 500 millions of tweets are posted every day on Twitter. Moreover, this data results from interactions of millions of users over time and therefore includes numerous complex structures. In this context, it is difficult for users to have a concrete vision of trends taking place and, even more, to apprehend the way in which all interactions are organised and can lead to media events.\\
In this paper, we seek to make this task achievable. More precisely, we aim at finding outliers in interaction data formed from a set of retweets. For instance, an event in a data stream is an outlier: it can be view as a statistical deviation of the total number of retweets at a given point in time. More generally, outliers, depending on which dimensions define them, highlight instants, users, users during given periods, or interactions for which the retweeting process behave unusually. Therefore, they constitute important information which is worth noticing from the perspective of the user. In order to find these unexpected behaviors, we design a multidimensional and multilevel analysis method.\\
First of all, we consider an interaction on Twitter to be a triplet $(s,a,t)$ meaning that user $s$, called the spreader, has retweeted a tweet of user $a$, called the author, at time $t$. We model the set of interactions as a data cube with three dimensions: spreaders, authors and time. This representation enables us to access local information, that is the number of retweets between two users during a specific hour, as well as more global and aggregated information, as for instance, the total number of retweets during a given hour. Afterwards, we combine and compare these different quantities of interactions between them in order to find outliers according to different contexts. Using the two previous quantities, we could, for instance, find an unexpected relationship between a spreader $s$ and an author $a$ during an hour $h$, if the number of retweet from $s$ to $a$ during $h$ is significantly large \textit{given} the total number of retweets observed during this hour. This analysis gives us insight into the possible reasons why some events emerge more than others and, in particular, whether they are global phenomena or, on the contrary, whether they originate from specific actors only.\\
Our method applies to all types of temporal interaction networks. One can add attributes to interactions by adding dimensions to the problem. In this paper, we add a semantic dimension referring to tweet contents by considering the $4$-tuples $(s,a,k,t)$ meaning that $s$ retweeted a tweet written by $a$ and containing hashtag $k$ at time $t$. This allows us to explore interactions from other perspectives and gain crucial information on events taking place.\\
The paper is organized as follows. We review the related work on outlier detection within Twitter in Section \ref{sec:Related Work}. We introduce the modelling of interactions as a data cube in Section \ref{sec:Data Cubes}, then we describe our method to build relevant contexts in Section \ref{sec:Our Method}. After describing our datasets in Section \ref{sec:dataset}, we present a case study in Section \ref{sec:Application to Political Communication on Twitter}. In particular, we investigate the causes of emergence of events found in the temporal dimension by exploring authors, spreaders, then hashtags dimensions. In Section \ref{sec:discussion}, we discuss two future works that can be achieved using our method, in particular, a characterization of the second screen usage and a user-topic link prediction. Finally, we conclude the paper in Section \ref{sec:Conclusion}.
 
\section{Related Work}
\label{sec:Related Work}
The problem of outlier detection on Twitter has attracted a significant amount of interest among scientists and has been approached in various ways depending on how outliers are defined and on the techniques used.\\ 
Some researchers consider outliers as real-world events happening at a given place and at a given moment. For example, Sakaki et al. \cite{sakaki2010earthquake} and Bruns et al. \cite{bruns2012qldfloods} trace specific \textit{keywords} attributed to a real-world event and find such outliers by monitoring temporal changes in word usage within tweets. There are also methods based on tweet clustering. In these approaches, authors infer, from timestamps, geo-localizations and tweet contents, a similarity between each pair of tweet and find real-world events into clusters of similar tweets. These techniques include the one of Dong et al. \cite{dong2015multiscale}, which computes similarities with a wavelet-based method between time series of keywords; the one of Li et al. \cite{li2012tedas} which aims at finding crime and disaster related events in a real time fashion; and the one of Walther et al. \cite{walther2013geo} which focuses on small scale events located in space.\\
Other researchers, instead, seek entities like bots, spammers, hateful users or influential \textit{users}. Thus, they consider outliers as users with abnormal behaviors according to different criteria. Varol et al. \cite{varol2017online} detect bots by means of a supervised machine learning technique. They extract features related to user activities along time, user friendships as well as tweet contents and use these features to identify bots by means of a labelled dataset. Stieglitz et al. \cite{stieglitz2012political} identify influential users by investigating the correlation between the vocabulary they use in tweets and the number of time they are retweeted. Ribeiro et al. \cite{ribeiro2018characterizing} detect hateful users. They start by classifying users with a lexicon-based method and then show that hateful users differ from normal ones in terms of their activity patterns and network structure.\\
Finally, other works aim at finding privileged \textit{relationships} between users. Among those, the work of Wong et al. \cite{wong2016quantifying} apply it to political leaning by combining an analysis of the number of retweets between two users with a sentiment analysis on the retweeted tweets. \\ 
All these works, although providing meaningful results, use different methods for different kind of outliers. Moreover, they only consider one perspective in the way they define them. With our approach, we want to treat these different types of outliers -- \textit{keywords, users, relationships} -- in a unified way as well as to consider different contexts in which outliers are considered abnormal. Hence, not only we consider different entities as abnormal users; abnormal relationships; abnormal behaviors of users during specific hours, \textit{etc.}, but also different contexts in which outliers are defined. Thus, an abnormal user may be abnormal during a given hour compared to the way it usually behaves during other hours, but also compared to the behavior of all other users during the same hour. In this way, our framework aims to give a more complete and systematic picture of how users act, interact, and are organized along time in a way similar to what Grasland et al. \cite{grasland2016international} do in the case of media coverage in newspapers.\\
In practice, instead of characterizing and detecting outliers using tweets'content, as a lot of current approaches do, included those set out above, we focus on the volume and structure of interactions. Indeed, text-mining techniques face challenges as the ambiguity of the language and the fact that resultant models are language-dependent and topic-depen-dent. Moreover, the structure of communication alone is already quite informative. Other authors point into this direction. For instance Chavoshi et al. \cite{chavoshi2017temporal} use a similar technique to the one of Varol et al. \cite{varol2017online}, but only exploit user activities through their number of tweets and retweets. In the same idea, Chierichetti et al. \cite{chierichetti2014event} look at the tweet/retweet volume and detect points in time when important events happen. Instead of volume-based features, another alternative to text-mining techniques is to use graph-based features. Song et al. \cite{song2011spam}, for instance, identify spammers in real time with a measure of distance and connectivity between users in the directed friendship graph (followers and followees). Bild et al. \cite{bild2015aggregate} designed a similar method but based on the retweet graph instead. Also based on the retweet graph, the method of Ten et al. \cite{ten2014modelling} detects trends by noticing changes in the size and in the density of the largest connected component. Another example is the approach of Coletto et al. \cite{coletto2017automatic} which combines an analysis of the friendship graph and of the retweet graph to identify controversies in threads of discussion.\\ 
In this paper, we design a method able to handle multiple types of outliers by observing the retweets' volume in numerous different contexts. We believe that this multidimensional and multilevel analysis is essential to detect subtle unexpected behaviors as well as fully understand the way in which millions of interactions may result in media events. 

\section{Formalism} 
\label{sec:Data Cubes}
We denote the set of interactions by a set $E$ of triplets such that $(s,a,t) \in E$ indicates that user $s$, called the spreader, has retweeted a tweet written by user $a$, called the author, at time $t$. We represent this set of interactions by a data cube \cite{han2011data}. In this section, we formally define this tool as well as the possible operations we can perform to manipulate data.

\subsection{Data Cube Definition}
A data cube is a general term used to refer to a multidimensional array of values \cite{han2011data}. Given $N$ dimensions characterized by $N$ sets $X_1,...,X_N$, we can built $\sum_{i=0}^{N}\binom{N}{N-i}$ data cubes, each representing a different degree of aggregation of data. The quantity $\binom{N}{N-i}$ corresponds to the number of data cubes of dimension $N-i$ in which $i$ dimensions are aggregated. Within this set of data cubes, we call the base cuboid $\mathcal{C}_{base}$ the $N$-dimensional data cube which has the lowest degree of aggregation. More generally, a $n$-dimensional data cube is denoted $\mathcal{C}_n(X,f)$ where $X=X_1\times ... \times X_n$ is the Cartesian product of the $n$ sets $X_1,...,X_n$, and $f$ is a feature which maps each $n$-tuple to a value in a value space $W$:
$$f :\quad X \longrightarrow W \quad\quad \quad \:$$
\vspace*{-0.7cm}$$ (x_1,...,x_n)  \longmapsto  f(x_1,...,x_n)\;.$$
In the following, $n$-tuples are also called cells of the cube and denoted $x$ such that $x=(x_1,...,x_n)\in X$. \\

\textit{Dimensions} are the sets of entities with respect to which we want to study data. As a first step, we can consider three dimensions: spreaders, denoted $S$, authors, denoted $A$, and time, denoted $T$. In addition, we can organise elements of a dimension into sub-dimensions. For instance, the temporal dimension can be organised depending on temporal granularity. In our case, we divide it into the two sub-dimensions days, denoted $D$, and hours of the day, denoted $H$, such that $t=(d,h)$ denotes the hour $h$ of day $d$, with $(d,h)\in D\times H$. While the set of days $D$ depends on the dataset, $H$ is the set of hours of the day such that $H=\{0,\cdots,23\}$. \\

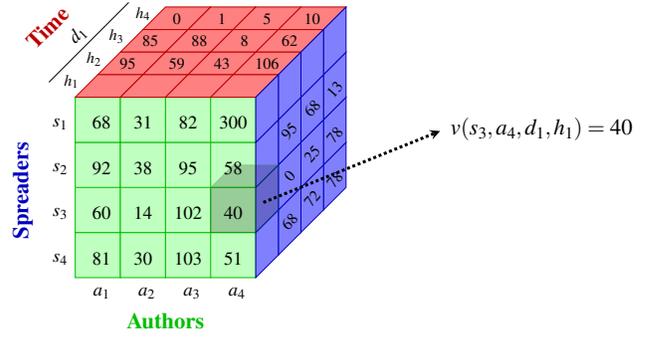
\begin{figure}
\scalebox{0.5}{
\begin{tikzpicture}
    \tikzcuboid{%
        shiftx=0cm,%
        shifty=0cm,%
        scale=1.2,%
        rotation=0,%
        densityx=1,%
        densityy=1,%
        densityz=1,%
        dimx=4,%
        dimy=4,%
        dimz=4,%
        linefront=green!75!black,%
        linetop=red!75!black,%
        lineright=blue!75!black,%
        fillfront=green!25!white,%
        filltop=red!50!white,%
        fillright=blue!50!white,%
    }
    
      \begin{scope}[transparency group, opacity=0.20]

     \tikzcuboid{%
        shiftx=1.8cm,%
        shifty=-0.6cm,%
        scale=1.2,%
        rotation=0,%
        densityx=1,%
        densityy=1,%
        densityz=1,%
        dimx=1,%
        dimy=1,%
        dimz=1,%
        linefront=green!75!black,%
        linetop=green!75!black,%
        lineright=green!75!black,%
        fillfront=black,%
        filltop=black,%
        fillright=black,%
    }

    \end{scope}    
  
       \draw[-stealth,line width=0.08cm, dotted] (2.6,-0.4) -- (7.3,1.5); 
    \node[label={\LARGE{$v(s_3,a_4,d_1,h_1)=40$}}] at (10,1) {}; 
    \node[label={[rotate=45]\textbf{\LARGE{\textcolor{red!75!black}{Time}}}}] at (-2.9,3.9) {};

    \node[label={[rotate=90]\textbf{\LARGE{\textcolor{blue!75!black}{Spreaders}}}}] at (-3.4,-0.2) {};
    \node[label={\textbf{\LARGE{\textcolor{green!75!black}{Authors}}}}] at (0,-4) {};


\node[label={\Large{$a_1$}}] at (-1.7,-3.2) {};
\node[label={\Large{$a_2$}}] at (-0.5,-3.2) {};     
\node[label={\Large{$a_3$}}] at (0.7,-3.2) {};
\node[label={\Large{$a_4$}}] at (1.9,-3.2) {};

    \node[label={\Large{60}}] at (-1.7,-1.1) {};
	\node[label={\Large{68}}] at (-1.7,1.3) {};
	\node[label={\Large{81}}] at (-1.7,-2.3) {};
	\node[label={\Large{92}}] at (-1.7,0.1) {};
	the set of hours
	\node[label={\Large{82}}] at (0.6,1.3) {};
	\node[label={\Large{95}}] at (0.6,0.1) {};
	\node[label={\Large{102}}] at (0.6,-1.1) {};
	\node[label={\Large{103}}] at (0.6,-2.3) {};
	
	\node[label={\Large{14}}] at (-0.6,-1.1) {};
	\node[label={\Large{31}}] at (-0.6,1.3) {};
	\node[label={\Large{30}}] at (-0.6,-2.3) {};
	\node[label={\Large{38}}] at (-0.6,0.1) {};
	
	\node[label={\Large{40}}] at (1.8,-1.1) {};
	\node[label={\Large{300}}] at (1.8,1.3) {};
	\node[label={\Large{51}}] at (1.8,-2.3) {};
	\node[label={\Large{58}}] at (1.8,0.1) {};

\node[label={\Large{$s_1$}}] at (-2.8,1.3) {};   
\node[label={\Large{$s_2$}}] at (-2.8,0.1) {};
\node[label={\Large{$s_3$}}] at (-2.8,-1.1) {};
\node[label={\Large{$s_4$}}] at (-2.8,-2.3) {};

\node[label={[rotate=45]\Large{$d_1$}}] at (-2.1,3.7) {}; 
    \draw (-3.1,2.8) -- +(45:3cm);
\node[label={\large{$h_1$}}] at (-2.5,2.4) {};  
\node[label={\large{$h_2$}}] at (-1.9,3) {};  
\node[label={\large{$h_3$}}] at (-1.3,3.6) {};
\node[label={\large{$h_4$}}] at (-0.6,4.2) {};

\node[label={\large{95}}] at (-1,2.9) {};
\node[label={\large{59}}] at (0.3,2.9) {};
\node[label={\large{43}}] at (1.5,2.9) {};
\node[label={\large{106}}] at (2.7,2.9) {};

\node[label={\large{85}}] at (-0.4,3.5) {};
\node[label={\large{88}}] at (0.9,3.5) {};
\node[label={\large{8}}] at (2.1,3.5) {};
\node[label={\large{62}}] at (3.3,3.5) {};

\node[label={\large{0}}] at (0.3,4.1) {};
\node[label={\large{1}}] at (1.5,4.1) {};
\node[label={\large{5}}] at (2.7,4.1) {};
\node[label={\large{10}}] at (3.9,4.1) {};

\node[label={[rotate=48]\large{68}}] at (3.5,-1.2) {};
\node[label={[rotate=48]\large{72}}] at (4.1,-0.6) {};
\node[label={[rotate=48]\large{78}}] at (4.7,-0.1) {};

\node[label={[rotate=48]\large{0}}] at (3.5,0) {};
\node[label={[rotate=48]\large{25}}] at (4.1,0.6) {};
\node[label={[rotate=48]\large{78}}] at (4.7,1.1) {};

\node[label={[rotate=48]\large{95}}] at (3.5,1.2) {};
\node[label={[rotate=48]\large{68}}] at (4.1,1.8) {};
\node[label={[rotate=48]\large{13}}] at (4.7,2.3) {};

\end{tikzpicture}
 
}
\caption{\small{\textbf{Base Cuboid $\mathcal{C}_4(S\times A \times D \times H,v)$ --} The base cuboid is not aggregated along any of its dimensions. It contains local information with respect to the quantity of interaction $v$. For instance, the gray cell indicates that $s_3$ retweeted $a_4$ 40 times on day $d_1$ at hour $h_1$.}}
\label{fig:base_cuboid}
\end{figure}

\textit{The feature} is a numerical measure which provides the quantities according to which we want to analyse relationships between dimensions. Here, we consider the quantity of interaction, denoted $v$. It gives the number of retweets for any combination of the four dimensions. In the base cuboid $\mathcal{C}_{base}=\mathcal{C}_4(S\times A\times D\times H,v)$, $v$ gives the number of times $s$ retweeted $a$ during hour $h$ of day $d$ (see Figure \ref{fig:base_cuboid}):
$$v :\:\: S\times A \times D \times H \longrightarrow  \mathbb{N}\:.$$

Data cubes of smaller dimensions are obtained by aggregating the base cuboid along one or several dimensions. We discuss this operation along with others in the next subsection.

\subsection{Data Cube Operations}

We can explore the data through three operations called aggregation, expansion and filtering.\\

\textit{Aggregation} is the operation which consists in seeing information at a more global level. Given a data cube $\mathcal{C}_n(X,f)$, the aggregation operation along dimension $X_i$ leads to a data cube of dimension $n-1$, $\mathcal{C}_{n-1}(X',f)$ where $X'=X_1 \times ... \times X_{i-1} \times X_{i+1} \times ... \times X_{n}$. Formally, a dimension $X_i$ is aggregated by adding up values of the feature for all elements $x_i\in X_i$. We indicate by a $\cdot$ the dimension which is aggregated with respect to $f$. Hence, $\mathcal{C}_{n-1}(X',f)$ is constituted of $n-1$-dimensional cells denoted $x'=(x_1,...,x_{i-1},\cdot,x_{i+1},...,x_{n})\in X'$ where
$$f(x')=\sum_{x_i\in X_i} f(x)\quad .$$
For instance, one can aggregate along the dimension of hours of the day such that
$$v(s,a,d,\cdot)=\sum_{h\in H} v(s,a,d,h)$$
gives the total number of time $s$ retweeted $a$ during day $d$. Opposed to the base cuboid, the apex cuboid is the most summarized cuboid. It is aggregated along all dimensions and hence consists in only one cell containing the grand total $f(\cdot,...,\cdot)=\sum_{x_1\in X_1}...\sum_{x_n\in X_n} f(x)$. In our case, the apex cuboid contains the total number of retweets. \\

We can also aggregate interactions according to a set of subsets of dimension $X_i$. Let $P_i$ denote this partition such that the intersection of any two distinct sets in $P_i$ is empty and the union of the sets in $P_i$ is equal to $X_i$. Then, given a data cube $\mathcal{C}_n(X,f)$, the aggregation operation along $P_i$ leads to a data cube $\mathcal{C}_n(X',f)$ with $X'=X_1\times ... \times X_{i-1}\times P_i \times X_{i+1}\times ... \times X_n$. This cube is constituted of $n$-dimensional cells denoted $x=(x_1,...,x_{i-1},C_k,x_{i+1},...,x_{n})\in X'$, with $C_k\in P_i$, such that 
$$f(x')=\sum_{x_i\in C_k} f(x)\quad .$$
For instance, one can aggregate according to a partition of hours $P_H=\{H_N,H_D\}$, where $H_N$ is the set of nocturnal hours and $H_D$ the set of daytime hours such that
$$v(s,a,d,H_N)=\sum_{h'\in H_N} v(s,a,d,h')$$
in $\mathcal{C}_4(S\times A \times D \times P_H)$, gives the total number of time $s$ retweeted $a$ during nocturnal hours on day $d$. \\

\textit{Expansion} is the reverse operation which consists in seeing information at a more local level by introducing additional dimensions. Given a data cube $\mathcal{C}_n(X,f)$, the expansion operation on dimension $X_{n+1}$ leads to a data cube of dimension $n+1$, $\mathcal{C}_{n+1}(X',f)$ where $X'=X \times X_{n+1} $.\\

\textit{Filtering} is the operation which consists in focusing on one specific subset of data. Given a data cube $\mathcal{C}_n(X,f)$, the filtering operation leads to a sub-cube $\mathcal{C}_n(X',f)$ by selecting subsets of elements within one or more dimensions such that $X'=X'_1 \times ... \times X'_n$ with $X'_1\subseteq X_1, ..., X'_n\subseteq X_n$.\\

It is also possible to combine operations together. For instance, we can filter the data cube aggregated on the partition of hours, $\mathcal{C}_4(S\times A \times D\times P_H,v)$, in order to focus on spreaders that abnormally retweet authors overnight on a given day. Note that the resulting data cube $\mathcal{C}_4(S\times A \times D\times \{H_N\},v)$ is different from the data cube  $\mathcal{C}_4(S\times A \times D\times H_N,v)$: in the first case, a cell $(s,a,d,H_N)$ gives the total number of time $s$ retweeted $a$ during nocturnal hours on day $d$; while in the second case, a cell $(s,a,d,h)$ give the number of times $s$ retweeted $a$ during hour $(d,h)$ where $h\in H_N$ is a nocturnal hour.\\

\begin{figure*}
  \begin{minipage}[c]{0.5\textwidth}
\scalebox{.95}{
\begin{tikzpicture}
\node[label={\textbf{Base cuboid}}] at (0.3,1) {}; 
\node[label={\textbf{Apex cuboid}}] at (0,-9) {};

  

\node[label={\textbf{Aggregation}}] at (-2.5,-1.7) {}; 
 \node[label={\textbf{on \textit{authors}}}] at (-2.5,-2) {};  
  
  \node[label={\textbf{Expansion}}] at (-3.2,-4.7) {}; 
 \node[label={\textbf{on \textit{time}}}] at (-3.2,-5) {};  
 
    \tikzcuboid{%
        shiftx=-3.5cm,%
        shifty=3.3cm,%
        scale=0.2,%
        rotation=0,%
        densityx=1,%
        densityy=1,%
        densityz=1,%
        dimx=2,%
        dimy=1,%
        dimz=4,%
        linefront=green!75!black,%
        linetop=red!75!black,%
        lineright=black,%
        fillfront=green!25!white,%
        filltop=red!50!white,%
        fillright=gray,%
    }
\node[label={(\textit{\small{spreaders, authors, time}})}] at (0,-1.1) {};

    \tikzcuboid{%
        shiftx=-3.8cm,%
        shifty=1cm,
        scale=0.2,%
        rotation=0,%
        densityx=1,%
        densityy=1,%
        densityz=1,%
        dimx=2,%
        dimy=4,%
        dimz=4,%
        linefront=green!75!black,%
        linetop=red!75!black,%
        lineright=blue!75!black,%
        fillfront=green!25!white,%
        filltop=red!50!white,%
        fillright=blue!50!white,%
    }
 

\draw[-stealth,line width=0.05cm,dotted] (2.25,0.7) to (2.8,1.2);
\draw[-,line width=0.05cm,dotted] (1,-0.3) to (1.4,0.025); 
\node[label={\textbf{Aggregation on}}] at (2.3,0.15) {}; 
\node[label={\textbf{\textit{authors partition}}}] at (2.3,-0.15) {}; 

    \tikzcuboid{%
        shiftx=3cm,%
        shifty=1.8cm,%
        scale=0.2,%
        rotation=0,%
        densityx=1,%
        densityy=1,%
        densityz=1,%
        dimx=4,%
        dimy=2,%
        dimz=4,%
        linefront=green!60!black,%
        linetop=red!75!black,%
        lineright=blue!75!black,%
        fillfront=green!80!black,%
        filltop=red!50!white,%
        fillright=blue!50!white,%
    }

\draw[-,line width=0.05cm,dotted] (-0.6,-0.3) to (-1.2,0.025); 
\draw[-stealth,line width=0.05cm,dotted] (-2.6,0.8) to (-3.3,1.2);
\draw[-stealth,line width=0.05cm] (-3.6,2) to (-3.6,2.8);
\node[label={\textbf{Aggregation}}] at (-2.7,2.1) {}; 
 \node[label={\textbf{on \textit{time}}}] at (-2.7,1.85) {};
   \node[label={\textbf{Filtering}}] at (-1.9,0.15) {}; 
 \node[label={\textbf{on \textit{spreaders}}}] at (-1.9,-0.15) {};  

    \tikzcuboid{%
        shiftx=0cm,%
        shifty=0cm,%
        scale=0.2,%
        rotation=0,%
        densityx=1,%
        densityy=1,%
        densityz=1,%
        dimx=4,%
        dimy=4,%
        dimz=4,%
        linefront=green!75!black,%
        linetop=red!75!black,%
        lineright=blue!75!black,%
        fillfront=green!25!white,%
        filltop=red!50!white,%
        fillright=blue!50!white,%
    }
\node[label={(\textit{\small{spreaders, authors, time}})}] at (0,-1.1) {};    
    
   \tikzcuboid{%
        shiftx=-2.5cm,%
        shifty=-3cm,%
        scale=0.2,%
        rotation=0,%
        densityx=1,%
        densityy=1,%
        densityz=1,%
        dimx=4,%
        dimy=4,%
        dimz=1,%
        linefront=green!75!black,%
        linetop=black,%
        lineright=blue!75!black,%
        fillfront=green!25!white,%
        filltop=gray,%
        fillright=blue!50!white,%
    }
\node[label={(\textit{\small{spreaders, time}})}] at (-2.5,-3.8) {};
    
       \tikzcuboid{%
        shiftx=0cm,%
        shifty=-2.5cm,%
        scale=0.2,%
        rotation=0,%
        densityx=1,%
        densityy=1,%
        densityz=1,%
        dimx=4,%
        dimy=1,%
        dimz=4,%
        linefront=green!75!black,%
        linetop=red!75!black,%
        lineright=black,%
        fillfront=green!25!white,%
        filltop=red!50!white,%
        fillright=gray,%
    }
\node[label={(\textit{\small{spreaders, authors}})}] at (0,-3.8) {};
    
       \tikzcuboid{%
        shiftx=2.5cm,%
        shifty=-2.5cm,%
        scale=0.2,%
        rotation=0,%
        densityx=1,%
        densityy=1,%
        densityz=1,%
        dimx=1,%
        dimy=4,%
        dimz=4,%
        linefront=black,%
        linetop=red!75!black,%
        lineright=blue!75!black,%
        fillfront=gray,%
        filltop=red!50!white,%
        fillright=blue!50!white,%
    }
\node[label={(\textit{\small{authors, time}})}] at (2.5,-3.8) {};    
    
       \tikzcuboid{%
        shiftx=-2.5cm,%
        shifty=-5.75cm,%
        scale=0.2,%
        rotation=0,%
        densityx=1,%
        densityy=1,%
        densityz=1,%
        dimx=4,%
        dimy=1,%
        dimz=1,%
        linefront=green!75!black,%
        linetop=black,%
        lineright=black,%
        fillfront=green!25!white,%
        filltop=gray,%
        fillright=gray,%
    }
\node[label={(\textit{\small{spreaders}})}] at (-2.5,-6.5) {}; 
    
       \tikzcuboid{%
        shiftx=0cm,%
        shifty=-5.75cm,%
        scale=0.2,%
        rotation=0,%
        densityx=1,%
        densityy=1,%
        densityz=1,%
        dimx=1,%
        dimy=4,%
        dimz=1,%
        linefront=black,%
        linetop=black,%
        lineright=blue!75!black,%
        fillfront=gray,%
        filltop=gray,%
        fillright=blue!50!white,%
    }
\node[label={(\textit{\small{time}})}] at (0,-6.5) {};  
    
       \tikzcuboid{%
        shiftx=2.5cm,%
        shifty=-5.5cm,%
        scale=0.2,%
        rotation=0,%
        densityx=1,%
        densityy=1,%
        densityz=1,%
        dimx=1,%
        dimy=1,%
        dimz=4,%
        linefront=black,%
        linetop=red!75!black,%
        lineright=black,%
        fillfront=gray,%
        filltop=red!50!white,%
        fillright=gray,%
    }
\node[label={(\textit{\small{authors}})}] at (2.5,-6.5) {};  
    
           \tikzcuboid{%
        shiftx=0cm,%
        shifty=-8cm,%
        scale=0.2,%
        rotation=0,%
        densityx=1,%
        densityy=1,%
        densityz=1,%
        dimx=1,%
        dimy=1,%
        dimz=1,%
        linefront=black,%
        linetop=black,%
        lineright=black,%
        fillfront=gray,%
        filltop=gray,%
        fillright=gray,%
    }
    
\draw[-stealth,line width=.05cm] (-0.75,-1) -- (-1.75,-2);
\draw[-,line width=.02cm,dotted] (0,-1) -- (0,-2);
\draw[-,line width=.02cm,dotted] (0.75,-1) -- (1.75,-2);
  
\draw[-,line width=.02cm,dotted] (0.5,-3.75) -- (2.125,-4.8);
\draw[-,line width=.02cm,dotted] (-0.5,-3.75) -- (-2,-4.8);
\draw[stealth-,line width=.05cm] (-2.125,-3.75) -- (-2.125,-4.8);
\draw[-,line width=.02cm,dotted] (2.375,-3.75) -- (2.375,-4.8);

\draw[-,line width=.02cm,dotted] (-2,-3.75) -- (0,-4.8);
\draw[-, line width=.02cm,dotted] (2,-3.75) -- (0.125,-4.8);

\draw[-,line width=.02cm,dotted] (0.125,-6.4) -- (0.125,-7.4);
\draw[-,line width=.02cm,dotted] (2,-6.4) -- (0.25,-7.4);
\draw[-,line width=.02cm,dotted] (-2,-6.4) -- (0,-7.4);



\end{tikzpicture}
}
\end{minipage}
  \begin{minipage}[c]{0.44\textwidth}
\caption{\textbf{Set of data cubes obtained by considering the three dimensions: spreaders, authors and time. \\\\} 
\small{Each data cube models interactions under a particular perspective: \\
-- we can move from one cuboid to another either by aggregation or expansion. \\
-- we can aggregate on a partition. In the example (top-right cube), we aggregate the base cuboid over authors communities. \\
-- we can focus on a particular subset by filtering a given data cube. In the example (top-left cube), we focus on two spreaders within the base cuboid.\\
-- we can combine operations and aggregate a filtered data cube. \\\\
From top to bottom, we have access to more and more aggregated information, for instance:\\
-- the $4D$ cell $x=(s_1,a_2, d_1,h_4)$ associated to the value $v(s_1,a_2,d_1,h_4)=9$ means that $s_1$ has retweeted $a_2$ 9 times on day $d_1$ during hour $h_4$;\\
-- the $3D$ cell $x=(a_2, d_1, h_4)$ associated to the value $v(\cdot,a_2,d_1,h_4)=1,288$ means that $a_2$ has been retweeted $1,288$ times on day $d_1$ during hour $h_4$ (by all spreaders);\\
-- the $1D$ cell $x=a_2$ associated to the value $v(\cdot,a_2,\cdot,\cdot)=29,362$ means that $a_2$ has been retweeted $29,362$ times (in the whole dataset);\\
-- the $0D$ cell $x=(\cdot,\cdot, \cdot,\cdot)$ associated to the value $v(\cdot,\cdot,\cdot,\cdot)=1,142,004$ means that the total number of retweets is equal to $1,142,004$.}
\vspace*{0.5cm}
}
\label{fig:data_cube}
\end{minipage}
\end{figure*}
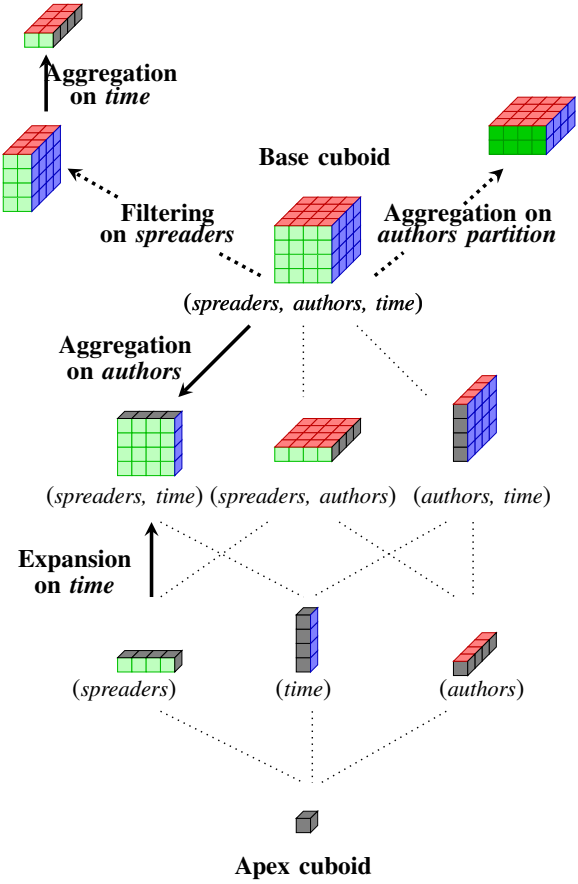

Figure \ref{fig:data_cube} shows a set of all data cubes that can be obtained considering the three dimensions: spreaders, authors and time. It also illustrates how to navigate from one to another thanks to the previously described operations.

\section{Method}
\label{sec:Our Method}
In this paper, our goal is to find abnormal data cube cells, \textit{i.e.}, entities $x\in X$ for which the observation $f(x)$ is abnormal. As an observation' abnormality is relative to the elements to which it is compared \cite{hawkins1980identification}, a given cell may be abnormal or not depending on the \textit{context}. The context, denoted $\mathscr{C}$, is the set of elements which are taken into account in order to assess the abnormality of an entity $x \in X$. In this section, we design a set of steps in order to shape various contexts and show, through several examples, that it leads to a deeper exploration of interactions compared to an elementary outlier detection.

\subsection{Construction of a Context}
An abnormal entity $x\in X$ is an entity which behavior deviates from its expected one. Hence, one way to find outliers in a set of entities $x\in X$ is to consider the following elements: \\
-- a set of observed values $\mathcal{O}=\{f(x), \, x\in X\}$;\\
-- a set of expected values $\mathcal{E}=\{f_{exp}(x), \, x\in X\}$;\\
-- a set of deviation values ${\mathcal{D}=\{d(f(x),f_{exp}(x)), \, x\in X\}}$, which quantify the differences between observed and expected values.\\
Together, these elements constitute the context $\mathscr{C}$. Then, given $\mathscr{C}$, an outlier $x\in X$ is a point whose absolute deviation value, $|d(f(x),f_{exp}(x))|$, is significantly larger than most others deviation values. \\

We build more or less elaborate contexts by playing with the considered observed, expected and deviation values.

\subsection{Observed values}
According to the type of unexpected behaviors we are looking for, the first step consists in choosing a cube among the set of cubes obtained from the base cuboid using one or several operations. This cube, denoted $\mathcal{C}_{obs}$, constitutes the set of entities and observed values.  \\

For instance, we can look for abnormal authors at given hours. To to so, we focus on the cube aggregated on spreaders such that $\mathcal{C}_{obs}=\mathcal{C}_3(A\times D\times H,v)$. We may also want to find abnormal authors during nocturnal hours only. In this case, we consider the aggregated and filtered data cube $\mathcal{C}_{obs}=\mathcal{C}_3(A \times D\times H_N,v)$.\\

In the first case, we consider all entities of the same type, $(a,d,h)\in A\times D\times H$: we are in a \textit{global context}. On the contrary, when we only consider a subset of all entities, as in the second example with $(a,d,h)\in A\times D\times H_N$, we are in a \textit{local context}.

\subsection{Expected values}

Once the set of observed values is fixed, we build a model of expected behavior based on a combination of other data cubes $\mathcal{C}_{m}(X',f)$, called \textit{comparison data cubes}. For the context to be relevant, these must derive from the aggregation of $\mathcal{C}_{obs}=\mathcal{C}_{n}(X,f)$ on one or more dimensions. Hence, $n>m$ and $X=X'\times Y$ where $Y$ is the Cartesian product of the aggregated dimensions. In the following, we build three different types of expected contexts: the basic, aggregative and multi-aggregative contexts. \\

\subsubsection{Basic Contexts}
When seeking abnormal cells within a data cube $\mathcal{C}_n(X,f)$, the most elementary context we can consider is the one in which the expected value is a constant, identical for each cell. We call it the \textit{basic context}. The model of expected behavior is that interactions are uniformly distributed over cells. In this case, the comparison data cube is the apex cuboid $\mathcal{C}_{0}(\cdot,f)$ and the expected value is the average number of interactions per cell:
$$f_{exp}(x)=\frac{f(\cdot)}{|X|}\: .$$ 

\noindent\textit{For instance, in data cube $\mathcal{C}_3(A\times D \times H, v)$, an abnormal cell $c^*=(a^*, d^*, h^*)$ indicates that during hour $h^*$ of day $d^*$, author $a^*$ has been retweeted an abnormal number of times compared to the average number of times any author is retweeted during any hour, $v_{exp}(a,d,h)=\frac{v(\cdot,\cdot,\cdot,\cdot)}{|A\times D \times H|}$.}

\subsubsection{Aggregative Contexts}
To find more subtle and local outliers, expected values must be more specific to each cell. The process is the same as in the basic context except that the considered comparison cube $\mathcal{C}_m(X',f)$ is not aggregated over all dimensions of $X$, \textit{i.e.} $X=X'\times Y$ with $Y\neq X$:
$$f_{exp}(x)=\frac{f(x')}{|Y|}\:,$$ 
such that $x=(x',y)\in X'\times Y$. Defined as such, the expected value is the value that one should observe if all interactions on $X'$ were homogeneously distributed on dimensions $Y$. We call these contexts, \textit{aggregative contexts}.\\

\noindent\textit{For instance, in data cube $\mathcal{C}_3(A\times D \times H, v)$, relatively to data cube $\mathcal{C}_2(D \times H,v)$ and expected values $$v_{exp}(a,d,h)=\frac{v(\cdot,\cdot, d,h)}{|A|}\,,$$
such that $Y=A$ and $X'=D\times H$, an abnormal cell $c^*=(a^*, d^*, h^*)$ indicates a significant deviation between the number of retweets received by $a^*$ during hour $(d^*,h^*)$ and the one that should have been observed if all authors had received the same number of retweets during hour $(d^*,h^*)$.}

\subsubsection{Multi-aggregative Contexts}
\label{subsec:Expected Context}
Aggregative contexts assume that interactions are homogeneously distributed among dimensions $Y$. It is possible to create contexts which differentiate the repartition of interactions according to each cell activity. We call them \textit{multi-aggregative contexts}. Unlike the other two, they require multiple comparison data cubes. There are no generic formulas: the number and types of comparison cubes as well as expected values depend on the application.\\

\noindent\textit{For instance, if we take back the previous example, we can consider, instead, the following expected values:
$$v_{exp}(a,d,h)=v(\cdot,\cdot, d,h)\times\frac{v(\cdot,a, \cdot,\cdot)}{v(\cdot,\cdot, \cdot,\cdot)}\,.$$
This way, it is expected that the number of retweets during $(d,h)$ is distributed among authors proportionally to their mean activity. We can also add information on authors activity during specific hours, and consider the cubes $\mathcal{C}_2(D \times H, v)$, $\mathcal{C}_2(A\times H,  v)$ and $\mathcal{C}_1(H,  v)$, such that
$$v_{exp}(a,d,h)=v(\cdot,\cdot, d,h)\times\frac{v(\cdot,a, \cdot,h)}{v(\cdot,\cdot, \cdot,h)}\,.$$
In this context, an abnormal cell $c^*=(a^*, d^*, h^*)$ indicates a significant deviation between the number of retweets received by $a^*$ during hour $h^*$ of day $d^*$ and the one that should have been observed if $a^*$ had been retweeted the way it is used to during hour $h^*$ on other days.\\ }

Each of these contexts can either be global or local depending on the chosen set of observed values within $\mathcal{C}_{obs}$.

\subsection{Deviation values}
Finally, for each cell $x$ within $\mathcal{C}_{obs}$, we measure the deviation between the observed value $f(x)$ and its expected value $f_{exp}(x)$. In this paper, we use two different deviation functions: the ratio and the Poisson deviation.\\

The \textit{ratio} between an observed value and an expected value is defined such that $$d_r(f(x),f_{exp}(x))=\frac{f(x)}{f_{exp}(x)}\:.$$
Note that this deviation function does not distinguish between $f(x)=2$ and $f_{exp}(x)=1$, on the one hand, and $f(x)=2,000$ and $f_{exp}(x)=1,000$, on the other hand.\\

To take into account the significance to which a value deviates, we define another deviation function: the \textit{Poisson deviation}. Indeed, in the cases in which the feature consists in counting the number of interactions during a given period, as $v(x)$, it can be modelled by a Poisson counting process of intensity $f_{exp}$ \cite{grasland2016international}, such that
$$\Pr(v(x)=k)=\frac{f_{exp}(x)^k e^{-f_{exp}(x)}}{k!}\:.$$
In this case, the Poisson deviation $d_p$ can be defined as follows. If $f(x)\leq f_{exp}(x)$, we calculate the probability of observing a value $f(x)$ or less, knowing that we should have observed $f_{exp}(x)$ on average. This probability is the cumulative distribution function of a Poisson distribution with parameter $f_{exp}(x)$. Accordingly, we denote it $F_{f_{exp}}(f(x))$. Then, by symmetry, we define $d_p$ such that:
\begin{equation*}
d_p(f(x),f_{exp}(x))=
\left\{
    \begin{array}{ll}
        \textcolor{white}{-}\log(F_{f_{exp}}(f(x))\:  \mbox{if } f(x)\leq f_{exp}(x),\\
        -\log(\bar{F}_{f_{exp}}(f(x)) \: \mbox{if } f(x)> f_{exp}(x),
    \end{array}
\right.
\label{eq:distance}
\end{equation*}
where the logarithm is calculated for convenience in order to have a better range of values. \\

In both cases, most of observed values are expected to be similar to their corresponding expected values. Consequently, the distribution of $\mathcal{D}$ is expected to follow a normal distribution in which most values fluctuates around a mean: $\bar{d_r}=1$ for the ratio and $\bar{d_p}=0$ for the Poisson deviation. Outlying cells, instead, correspond to deviation values significantly distant from the mean\footnote{We use the classical assumption that a value is anomalous if its distance to the mean exceeds three times the standard deviation \cite{chandola2009anomaly,han2011data}.}.\\

\subsection{Examples}

\begin{figure}[h]
\begin{tikzpicture}
\node[label={\scriptsize{$10,000$ retweets}}] at (-0.5,0.4) {};   
\pie[explode=0, before number=\phantom,after number=, rotate=-40, pos={-0.5,-0.5}, radius=1]{54/, 15/\small{${a_1}^\#$}, 12/\small{${a_2}^\#$} ,8/\small{$a_3$}, 4/\small{},4/\small{},3/\small{}};
\node[label={\scriptsize{$600$ retweets}}] at (2,0.4) {};   
\pie[explode=0, before number=\phantom,after number=, rotate=-40, pos={2,-0.5}, radius=0.4]{15/, 12/\small{${a_1}^\#$}, 50/\small{${a_2}^\#$} ,12/\small{$a_3$},  4/\small{},4/\small{},3/\small{}};
 \node[label={\Large{...}}] at (3.5,-1) {};   
 \node[label={\scriptsize{$600$ retweets}}] at (5.5,0.4) {}; 
\pie[explode=0, before number=\phantom,after number=, rotate=-40, pos={5.5,-0.5}, radius=0.4]{40/, 25/\small{${a_1}^\#$}, 15/\small{${a_2}^\#$} ,9/\small{$a_3$}, 4/\small{},4/\small{},3/\small{}};

\draw[-,line width=.05cm,dotted] (-1.5,1.15) -- (6.2,1.15);

\node[label={\small{$(d_1,19h)$}}] at (-0.5,1.1) {};    
\node[label={\small{$(d_2,19h)$}}] at (2,1.1) {};    
\node[label={\small{$(d_n,19h)$}}] at (5.5,1.1) {};     
    
  \node[label={\small{$\#$ influential author}}] at (5,-2.51) {};  
\draw [fill=blue!65] (1.1,-2.2) rectangle (1.5,-2);
   \node[label={\small{Other authors}}] at (2.5,-2.46) {}; 
  \end{tikzpicture} 
\caption{{\small \textbf{Different contexts lead to different outliers --} The numbers of retweets per hour distributed among authors are represented as pie charts. }}
\label{fig:proportions_auteurs}
\end{figure}
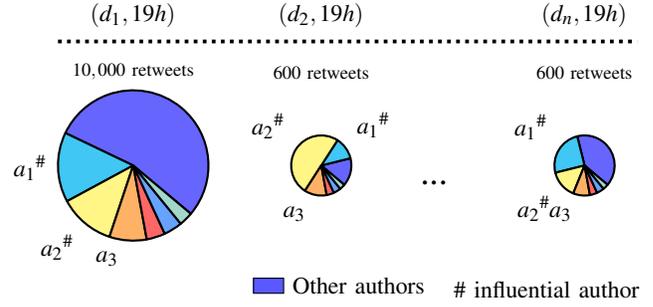

Figure \ref{fig:proportions_auteurs} illustrates several situations in which we find different abnormal authors during given hours by considering different contexts and a ratio deviation function:
-- Triplet $(a_1, d_1, 19h)$ is abnormal in the global basic context: it has been retweeted $1,500$ times ($15\%$ of a $10,000$) which is higher than all other triplets.\\
-- Triplet $(a_2, d_2, 19h)$ is abnormal in the global aggregative context: its proportion of retweet is $50\%$ which is higher than all other triplets.\\
-- Triplet $(a_1, d_n, 19h)$ is abnormal in the global multi-aggre-gative context: the deviation in the activity of $a_1$ with respect to its usual activity at $19h$ is higher than all other triplets.\\ 
-- Triplet $(a_3, d_2, 19h)$ is abnormal in the local aggregative context: its proportion of retweet is higher than other triplets $(a,d,h)$ in which $a$ is not an influential author.\\

As this example shows and as we will show in practice in the next sections, our approach, combining data cubes to build different contexts, leads to numerous kinds of outliers which allows us to thoroughly analyse temporal interactions under different perspectives.

\section{Datasets}
\label{sec:dataset}
In this paper, we choose to study the organization of interactions on Twitter by analysing different sets of politics-related retweets. Indeed, since Twitter is an integral part of means of communication used by political leaders to disseminate information to the public, finding abnormal entities corresponding to different kinds of unexpected behaviors in this situation is of great interest. To do so, we use two different datasets.\\

\textbf{Dataset $D_1$} is a set of retweets related to political communication during the 2017 French presidential elections. We use a subset of the dataset collected by Gaumont et al. \cite{gaumont2017methods} as part of the \textit{Politoscope} project. It contains politics-related retweets during the month of August 2016. Formally, our dataset consists in the set of retweets $E$, such that $(s,a,d,$ $h)\in E$ means that $s$ retweeted $a$ at hour $h$ of day $d$, where either the corresponding tweet contains politics-related keywords, or $a$ belongs to a set of $3, 700$ French political actors listed by the Politoscope project. It contains $1,142,004$ retweets and involves $211,155$ different users. In this dataset, the set of days is $D= \{1,\cdots,31\}$.\\

\textbf{Dataset $D_2$} is the same as dataset $D_1$ except that it contains an additional dimension. It consists in the set of re-tweets $E$, such that $(s,a,k,d,h)\in E$ means that $s$ retweeted a tweet written by $a$ and containing the hashtag $k$ at hour $h$ of day $d$. It contains $|K|=30,057$ different hashtags.\\

In the following, usernames are only mentioned when they correspond to official Twitter accounts of politicians, or public organizations, such as city halls, newspapers, or shows. Otherwise, they are designated by generic terms \textit{user-n}, where $n$ is an integer to differentiate anonymous users.

\section{Experiments}
\label{sec:Application to Political Communication on Twitter}

As a first illustration of our method, we present a case study which, based on events found in the temporal dimension, proposes possible causes of their emergence by exploring other dimensions. First, we apply our method on dataset $\mathcal{D}_1$ and focus on the three dimensions: spreaders, authors and time. Then, we add the hashtag dimension with dataset $\mathcal{D}_2$ in order to gain more insight on events.

\subsection{Events}
\label{subsec:hours}
We define an event $e=((d^*_1,h^*_1),\cdots,(d^*_n,h^*_n)) \in \mathcal{E}$ to be a set of consecutive abnormal hours. For convenience, we denote it $e=(d^*,h^*_1\text{ - }h^*_n)$ when all hours span over the same day $d^*$.\\

\begin{figure*}
  \begin{minipage}[c]{0.79\textwidth}
 \resizebox{1\textwidth}{!}{\input{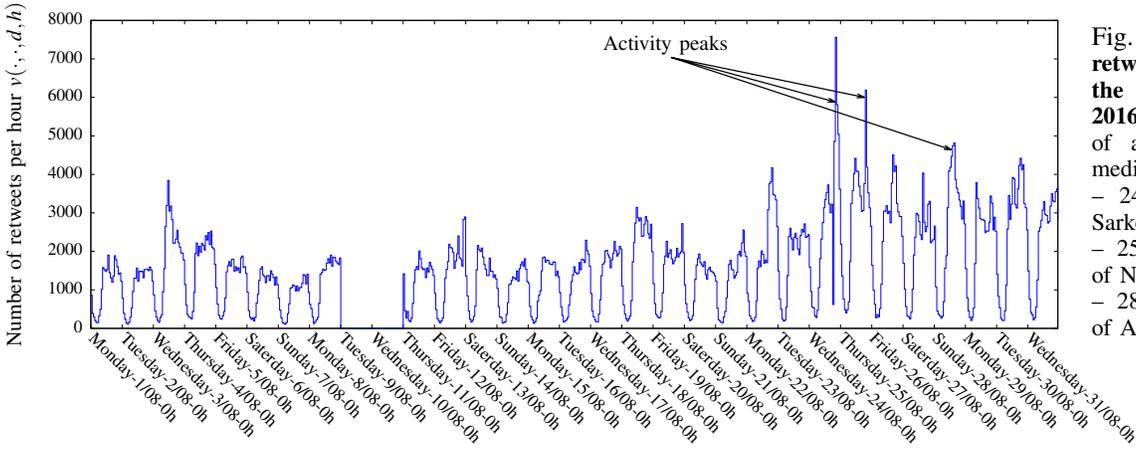}}
 \end{minipage}
  \begin{minipage}[c]{0.2\textwidth}
\caption{\small{\textbf{Number of retweets per hour along the month of August 2016 -- }The three peaks of activity correspond to media events:\\
-- $24/08$: interview of N. Sarkozy on television news,\\
-- $25/08$: political meeting of N. Sarkozy,\\
-- $28/08$: political meeting of A. Juppé.}}
\vspace*{1cm}
	\label{fig:evolution_retweets}
 \end{minipage}
\end{figure*}

Figure \ref{fig:evolution_retweets} shows the evolution of the number of retweets per hour\footnote{Note that due to a server failure from Tuesday the $9^{th}$ to Thursday the $11^{th}$, no activity is observed during this period.}. We can distinguish three distinct behaviors:\\
-- nocturnal hours, characterized by a number of retweets fluctuating \mbox{around $350$},\\
-- daytime from the $1^{st}$ of August to the $24^{th}$, characterized by a higher number of retweets fluctuating around $1,700$,\\
-- daytime from the $24^{th}$ of August to the $31^{st}$, characterized by a global increase in the number of retweets which fluctuates around $2,900$.

\subsubsection{Basic Context}
First of all, we look for events in the basic context. The sets of entities and observed values are provided by data cube $\mathcal{C}_2(D\times H,v)$. Expected values are defined such that 
$$v_{exp}^b(d,h)=\frac{v(\cdot,\cdot, \cdot,\cdot)}{|D\times H|}\: .$$
Figure \ref{fig:abnormal_hours} (Left) shows the distribution of deviation values by considering a ratio-based deviation. We find seven abnormal hours leading to three events such that $$\mathcal{E}=\{(24,20h\text{ - }22h), (25,19h), (28,14h\text{ - }15h)\}\:.$$ We see that these hours correspond to the three peaks of activity on Figure \ref{fig:evolution_retweets}. Hence, this context does not highlight local anomalies but only global ones, deviating from all observations. Therefore, it is biased by circadian and weekly rhythms and does not have access to abnormal nocturnal hours nor hours located during the first part of the month.

\subsubsection{Aggregative Context}
To take into account the overall increase in the number of retweets during the month, we need to use a aggregative context in which expected values incorporate the overall activity of the day provided by data cube $\mathcal{C}_1(D,v)$:
$$v_{exp}^a(d,h)=\frac{v(\cdot,\cdot,d,\cdot)}{|H|}\:.$$
As such, deviation values are independent of daily variations in the data. This is what we observe in Figure \ref{fig:abnormal_hours} (Center). We find 10 abnormal hours. Among those, six hours are part of the first period of the month: the $3^{rd}$ at $11h$, the $12^{th}$ at $23h$, the $21^{st}$ at $21h$, and the $22^{th}$ from $17h$ to $20h$. Nevertheless, extreme values are still biased by circadian rhythms which prevent us from detecting abnormal nocturnal hours.

\subsubsection{Multi-aggregative Context}
\begin{figure*}
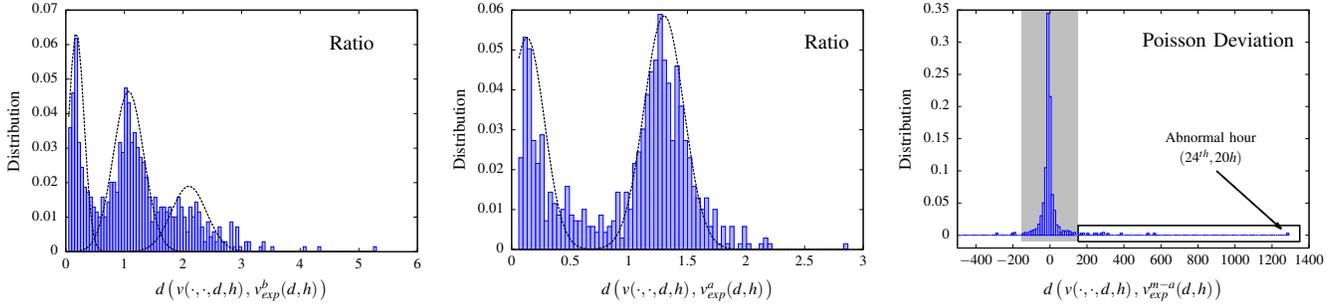

  \begin{minipage}[c]{0.32\textwidth}
 \resizebox{1\textwidth}{!}{\input{histo_retweets.tex}}
\end{minipage}
  \begin{minipage}[c]{0.32\textwidth}
 \resizebox{1\textwidth}{!}{\input{histo_retweets_proportion.tex}}
\end{minipage}
  \begin{minipage}[c]{0.32\textwidth}
 \resizebox{1\textwidth}{!}{\input{hours_aggregated_expected.tex}}
\end{minipage}
	\caption{\small{\textbf{Deviation values of hours in basic, aggregative, and multi-aggregative contexts -- }(Left) Basic -- The three bell curves correspond to three distinct behaviors: nocturnal hours ($\bar{d_r^1}=0.2$), daytime from the $1^{st}$ of August to the $24^{th}$ ($\bar{d_r^2}=1.1$) and daytime from the $24^{th}$ of August to the $31^{th}$ ($\bar{d_r^3}=2.1$). (Center) Aggregative -- We only observe two behaviors, the one corresponding to nocturnal hours which fluctuates around $\bar{d_r^1}=0.13$ and the one corresponding to daytime which fluctuates around $\bar{d_r^2}=1.3$. (Right) Multi-aggregative -- Most deviation values are centred on $\bar{d_p}=0$ (gray zone), meaning that they are likely to be generated by a Poisson counting process with intensity $v_{exp}^{m-a}(d,h)$. Values far away from the mean represent hours which behave significantly differently compared to the way they should.}}
	\label{fig:abnormal_hours}
\end{figure*}

To address this issue, we use a multi-aggregative context in which we add aggregated information relating to the typical activity per hour, provided by data cubes $\mathcal{C}_1(H,v)$ and $\mathcal{C}_0(\cdot,v)$:
$$v_{exp}^{m-a}(d,h)=v(\cdot,\cdot,d,\cdot)\times \frac{v(\cdot,\cdot,\cdot,h)}{v(\cdot,\cdot,\cdot,\cdot)}\:.$$
Moreover, we take the Poisson distance as a deviation measure to account for the significance of deviations. We find 40 abnormal hours (see Figure \ref{fig:abnormal_hours} (Right)). Among those, several are adjacent, which leads to 17 distinct events (see Table \ref{tab:events}). \\

Hour $(11^{th},0h)$ is abnormal. It means that, on average, at $0h$, we expect to observe $v(\cdot,\cdot,\cdot,0h)/v(\cdot,\cdot,\cdot,\cdot)=3.16\%$ of the total number of retweets of the day. Hence, on hour $(11^{th},0h)$, we expect to observe $v(\cdot,\cdot,11^{th},\cdot)\times 3.16\%=909$ retweets. However, we observe $1,418$ retweets in $\mathcal{C}_2(D\times H,v)$. This deviation from the expected value is much more important than those observed for most hours $(d,h)\in D\times H$. As a consequence, $(11^{th},0h)$ is an abnormal hour in this particular multi-aggregative context. \\

In Table \ref{tab:events}, we see several hours of generally low activity as nocturnal hours. This last result shows that using more sophisticated contexts leads to more subtle outliers. 


\begin{table*}
\setlength\tabcolsep{1.8pt}
\begin{center}
\begin{tabular}{|c|c|c|}
\hline
  \centered{ Events } &
  \centered{ Significant \\ Abnormal Authors} &  \centered{ Media Events}\\
         \hline
  \centered{$(3^{th},10h\text{ - }13h)$} & \centered{several}& \centered{Police intervention\\ in a church}\\
        \hline
  \centered{$(11^{th},0h)$} & \centered{marseille}& \centered{Fire in the city of marseille}\\
        \hline
  \centered{$(11^{th},3h)$} &  \centered{FrancoisFillon}& \centered{Unknown}\\
            \hline
  \centered{$((12^{th},22h),\cdots,$\\$(13^{th},1h))$} &  \centered{fhollande}& \centered{Olympic victory of France}\\
          \hline
  \centered{$(13^{th},9h)$} &  \centered{none}& \centered{Unknown}\\
            \hline
  \centered{$(19^{th},22h)$} &  \centered{none}& \centered{Olympic victory of France}\\
          \hline
  \centered{$(21^{th},21h)$} &  \centered{none}& \centered{Olympic victory of France}\\
          \hline
  \centered{$(22^{th},16h\text{ - }22h)$} &\centered{several}& \centered{Announcement of\\ N. Sarkozy's campaign}\\
        \hline
  \centered{$(23^{th},7h\text{ - }8h)$} &\centered{none}& \centered{Unknown}\\
        \hline
  \centered{$(24^{th},20h\text{ - }22h)$} &\centered{several}& \centered{Interview of N. Sarkozy \\on television news}\\
      \hline
  \centered{$(25^{th},19h)$} &\centered{NicolasSarkozy}& \centered{Political Meeting \\ of N. Sarkozy}\\
        \hline
  \centered{$(26^{th},15h\text{ - }18h)$} &\centered{several}& \centered{Council of state on \\ burkini wearing}\\
      \hline
  \centered{$(27^{th},15h)$} &\centered{alainjuppe}& \centered{Political Meeting \\ of A. Juppé}\\
        \hline
  \centered{$(28^{th},0h)$} &\centered{several}& \centered{Interview of N. Kosciusko-\\Morizet on a talk-show}\\
      \hline
  \centered{$(28^{th},13h\text{ - }15h)$} &\centered{JLMelenchon} & \centered{Political Meeting \\ of J-L. Mélenchon}\\
      \hline
  \centered{$(29^{th},7h\text{ - }9h)$} &\centered{NicolasSarkozy}& \centered{Interview of N. Sarkozy \\on a radio program}\\
    \hline
      \centered{$(30^{th},17h\text{ - }18h)$} &\centered{none}& \centered{Resignation of E. Macron\\ from government}\\
    \hline
\end{tabular}
\end{center}
\caption{\small{\textbf{List of detected abnormal events and authors together with their associated media events.}}}
\label{tab:events}
\end{table*}

\subsection{Abnormal authors during events}
\label{subsec:authors}
Now, we focus on determining whether an abnormal event is due to specific authors which have been retweeted predominantly, or, on the contrary, results from a more global phenomenon in which we observe an overall increase of the activity.\\

To do so, we use a local and multi-aggregative context. Observed values are provided by the filtered and aggregated data cube $\mathcal{C}_3(A\times \{e\},v)$, where $e\in \mathcal{E}$ is an abnormal event. A cell $(a,e)$ within this cube gives the total number of times author $a$ has been retweeted during event $e$. This way, we focus on how interactions are organized among authors within each event.\\

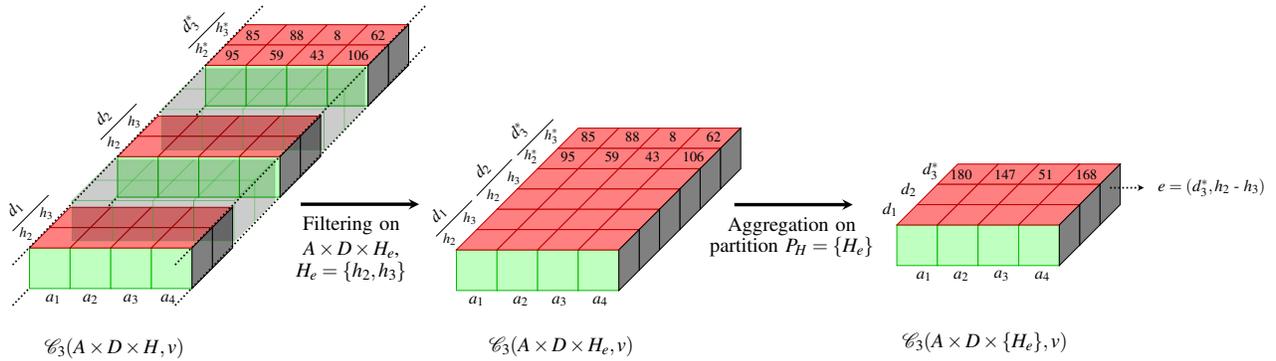
\begin{figure*}
\hspace*{.3cm}
\scalebox{0.45}{
\begin{tikzpicture}
	\node[label={\LARGE{$\mathcal{C}_3(A \times D \times H,v)$}}] at (0.7,-0.5) {};

        \draw[dotted,line width=0.05cm] (-2.3,1.3) -- +(46:10.7cm);
    \tikzcuboid{%
        shiftx=2cm,%
        shifty=5.7cm,%
        scale=1.2,%
        rotation=0,%
        densityx=1,%
        densityy=1,%
        densityz=1,%
        dimx=4,%
        dimy=1,%
        dimz=2,%
        linefront=green!75!black,%
        linetop=red!75!black,%
        lineright=black!75!black,%
        fillfront=green!25!white,%
        filltop=red!50!white,%
        fillright=black!50!white,%
    }

\node[label={[rotate=45]\Large{$d_2$}}] at (0.6,6.5) {}; 
\node[label={\large{$h_2$}}] at (0.7,5.7) {};  
\node[label={\large{$h_3$}}] at (1.3,6.3) {};
    \draw (0.3,6.2) -- +(45:1.5cm);

    
    \node[label={[rotate=45]\Large{$d_3^*$}}] at (3.2,9.2) {}; 
\node[label={\large{$h_2^*$}}] at (3.3,8.4) {};  
\node[label={\large{$h_3^*$}}] at (3.9,9) {};
    \draw (2.9,8.9) -- +(45:1.5cm);
    
        \tikzcuboid{%
        shiftx=4.6cm,%
        shifty=8.4cm,%
        scale=1.2,%
        rotation=0,%
        densityx=1,%
        densityy=1,%
        densityz=1,%
        dimx=4,%
        dimy=1,%
        dimz=2,%
        linefront=green!75!black,%
        linetop=red!75!black,%
        lineright=black!75!black,%
        fillfront=green!25!white,%
        filltop=red!50!white,%
        fillright=black!50!white,%
    }

    \tikzcuboid{%
        shiftx=-0.6cm,%
        shifty=3cm,%
        scale=1.2,%
        rotation=0,%
        densityx=1,%
        densityy=1,%
        densityz=1,%
        dimx=4,%
        dimy=1,%
        dimz=2,%
        linefront=green!75!black,%
        linetop=red!75!black,%
        lineright=black!75!black,%
        fillfront=green!25!white,%
        filltop=red!50!white,%
        fillright=black!50!white,%
    }
        \draw[dotted,line width=0.05cm] (-2.3,2.5) -- +(46:10.7cm); 
        \draw[dotted,line width=0.05cm] (2.5,2.5) -- +(46:10.7cm);       
        \draw[dotted,line width=0.05cm] (2.5,1.3) -- +(46:10.7cm);    

	\node[label={\Large{$a_1$}}] at (-1.1,1.1) {};
	\node[label={\Large{$a_3$}}] at (1.2,1.1) {};
	\node[label={\Large{$a_2$}}] at (0,1.1) {};
	\node[label={\Large{$a_4$}}] at (2.3,1.1) {};

\node[label={[rotate=45]\Large{$d_1$}}] at (-2,3.8) {}; 
\node[label={\large{$h_2$}}] at (-1.9,3) {};  
\node[label={\large{$h_3$}}] at (-1.3,3.6) {};
    \draw (-2.3,3.5) -- +(45:1.5cm);

\node[label={\large{95}}] at (4.2,8.3) {};
\node[label={\large{59}}] at (5.5,8.3) {};
\node[label={\large{43}}] at (6.7,8.3) {};
\node[label={\large{106}}] at (7.9,8.3) {};

\node[label={\large{85}}] at (4.8,8.9) {};
\node[label={\large{88}}] at (6.1,8.9) {};
\node[label={\large{8}}] at (7.3,8.9) {};
\node[label={\large{62}}] at (8.5,8.9) {};

      \begin{scope}[transparency group, opacity=0.20]

     \tikzcuboid{%
        shiftx=0.65cm,%
        shifty=4.4cm,%
        scale=1.2,%
        rotation=0,%
        densityx=1,%
        densityy=1,%
        densityz=1,%
        dimx=4,%
        dimy=1,%
        dimz=2,%
        linefront=green!75!black,%
        linetop=green!75!black,%
        lineright=green!75!black,%
        fillfront=black,%
        filltop=black,%
        fillright=black,%
    }

    \end{scope} 
    
          \begin{scope}[transparency group, opacity=0.20]

     \tikzcuboid{%
        shiftx=3.3cm,%
        shifty=7.1cm,%
        scale=1.2,%
        rotation=0,%
        densityx=1,%
        densityy=1,%
        densityz=1,%
        dimx=4,%
        dimy=1,%
        dimz=2,%
        linefront=green!75!black,%
        linetop=green!75!black,%
        lineright=green!75!black,%
        fillfront=black,%
        filltop=black,%
        fillright=black,%
    }

    \end{scope} 

\end{tikzpicture}
 
}

\vspace*{-3.4cm}
\hspace*{4.2cm}
\scalebox{0.45}{
\begin{tikzpicture}
	\node[label={\LARGE{$\mathcal{C}_3(A \times D \times H_e,v)$}}] at (1,-2.3) {};

	\node[label={\LARGE{Filtering on}}] at (-5.2,1.3) {};
		\node[label={\LARGE{$A\times D\times H_e$,}}] at (-5.2,0.6) {};
				\node[label={\LARGE{$H_e=\{h_2,h_3\}$}}] at (-5.2,-0.1) {};
\draw[-stealth,line width=0.1cm] (-6.7,2.45) to (-3.2,2.45);

    \tikzcuboid{%
        shiftx=1.5cm,%
        shifty=3.5cm,%
        scale=1.2,%
        rotation=0,%
        densityx=1,%
        densityy=1,%
        densityz=1,%
        dimx=4,%
        dimy=1,%
        dimz=6,%
        linefront=green!75!black,%
        linetop=red!75!black,%
        lineright=black!75!black,%
        fillfront=green!25!white,%
        filltop=red!50!white,%
        fillright=black!50!white,%
    }
    

	\node[label={\Large{$a_1$}}] at (-1.5,-0.8) {};
	\node[label={\Large{$a_3$}}] at (0.95,-0.8) {};
	\node[label={\Large{$a_2$}}] at (-0.2,-0.8) {};
	\node[label={\Large{$a_4$}}] at (2.15,-0.8) {};

\node[label={[rotate=45]\Large{$d_3^*$}}] at (0,4.2) {}; 
\node[label={\large{$h_2^*$}}] at (0.1,3.4) {};  
\node[label={\large{$h_3^*$}}] at (0.7,4) {};
    \draw (-0.3,3.9) -- +(45:1.5cm);
    
\node[label={[rotate=45]\Large{$d_2$}}] at (-1.1,3.1) {}; 
\node[label={\large{$h_2$}}] at (-1,2.3) {};  
\node[label={\large{$h_3$}}] at (-0.4,2.8) {};
    \draw (-1.5,2.7) -- +(46:1.5cm);
    
\node[label={[rotate=45]\Large{$d_1$}}] at (-2.4,1.8) {}; 
\node[label={\large{$h_2$}}] at (-2.3,1) {};  
\node[label={\large{$h_3$}}] at (-1.7,1.6) {};
    \draw (-2.7,1.5) -- +(45:1.5cm);

\node[label={\large{95}}] at (1.2,3.45) {};
\node[label={\large{59}}] at (2.5,3.45) {};
\node[label={\large{43}}] at (3.7,3.45) {};
\node[label={\large{106}}] at (4.9,3.45) {};

\node[label={\large{85}}] at (1.8,4.05) {};
\node[label={\large{88}}] at (3.1,4.05) {};
\node[label={\large{8}}] at (4.3,4.05) {};
\node[label={\large{62}}] at (5.5,4.05) {};

\end{tikzpicture}
 
}

\vspace*{-2.8cm}
\hspace*{9.7cm}
\scalebox{0.45}{
\begin{tikzpicture}
	\node[label={\LARGE{$\mathcal{C}_3(A \times D \times \{H_e\},v)$}}] at (1.3,-1.6) {};
	\node[label={\LARGE{Aggregation on}}] at (-4.4,1.85) {};
		\node[label={\LARGE{partition $P_{H}=\{H_e\}$}}] at (-4.4,1.2) {};
\draw[-stealth,line width=0.1cm] (-6.1,3) to (-2.6,3);
       \tikzcuboid{%
        shiftx=0.5cm,%
        shifty=3cm,%
        scale=1.2,%
        rotation=0,%
        densityx=1,%
        densityy=1,%
        densityz=1,%
        dimx=4,%
        dimy=1,%
        dimz=3,%
        linefront=green!75!black,%
        linetop=red!75!black,%
        lineright=black!75!black,%
        fillfront=green!25!white,%
        filltop=red!50!white,%
        fillright=black!50!white,%
    }
    

	\node[label={\Large{$a_1$}}] at (-0.5,.5) {};
	\node[label={\Large{$a_3$}}] at (1.8,.5) {};
	\node[label={\Large{$a_2$}}] at (0.6,.5) {};
	\node[label={\Large{$a_4$}}] at (3,0.5) {};

\node[label={\Large{$e=(d_3^*,h_2\text{ - }h_3)$}}] at (8,3) {}; 
\draw[-stealth,line width=0.05cm,dotted] (5,3.5) to (6,3.5);

\node[label={\Large{$d^*_3$}}] at (-0.3,3.5) {}; 
\node[label={\Large{$d_2$}}] at (-0.95,3) {}; 
\node[label={\Large{$d_1$}}] at (-1.5,2.4) {}; 

\node[label={\large{180}}] at (0.6,3.5) {};
\node[label={\large{147}}] at (1.9,3.5) {};
\node[label={\large{51}}] at (3.1,3.5) {};
\node[label={\large{168}}] at (4.3,3.5) {};

\end{tikzpicture}
}

\caption{\small{\textbf{Local and multi-aggregative context to focus on authors during events --} To investigate the possible causes of the emergence of event $e=(d^*_3,h_2^*\text{ - }h_3^*)$, we characterize the authors' usual behaviors during the corresponding time periods on other days. }}
\label{fig:schema_event}
\end{figure*}

We proceed in a similar way to obtain expected values. Instead of considering the set of authors during event $e$, we consider the set of authors during each of the hourly periods corresponding to $e$ on all days. We denoted this set of hours $H_e=\{h^*\in H \,|\, (d^*,h^*)\in e\}$. We focus on data cube $\mathcal{C}_3(A\times D\times P_{H},v)$, aggregated on the partition of $H$, $P_{H}=\{H_e\}$. Operations performed to switch from the original cube $\mathcal{C}_3(A\times D\times H,v)$ to data cube $\mathcal{C}_3(A\times D\times \{H_e\},v)$ is depicted in Figure \ref{fig:schema_event}. \\

Finally, expected values are defined using the comparison data cubes $\mathcal{C}_2(A\times \{H_e\},v)$ and $\mathcal{C}_1(\{H_e\},v)$, obtained by aggregation of $\mathcal{C}_3(A\times D\times \{H_e\},v)$, and data cube $\mathcal{C}_2(\{e\},v)$, obtained by aggregation and filtering of $\mathcal{C}_3(A\times D\times \{H_e\},v)$:
$$v_{exp}(a,e)=v(\cdot,\cdot,e) \times \frac{v(\cdot,a,\cdot,H_e)}{v(\cdot,\cdot,\cdot,H_e)}\: ,$$
where $v(\cdot,\cdot,e)=\sum_{(d^*,h^*)\in e} v(\cdot,\cdot,d^*,h^*)$, is the number of retweets observed during $e$; $v(\cdot,a,\cdot,H_e)$ is the total number of retweets author $a$ received during hours of $H_e$; and $v(\cdot,\cdot,\cdot,H_e)$ is the total number of retweets observed during $H_e$. \\

According to this context, a couple $(a^*,e)\in A\times \{e\}$ is abnormal when there is a significant deviation between the number of retweets received by $a$ during $e$, and the number of retweets $a$ is expected to receive on average during the corresponding period on other days. In the following, we discuss the three different situations which arise through specific examples. \\

\begin{figure*}
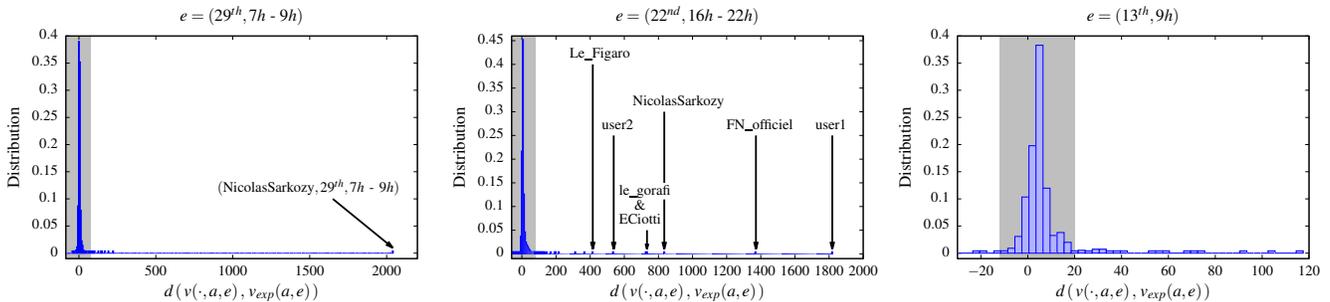

  \begin{minipage}[c]{0.32\textwidth}
 \resizebox{1\textwidth}{!}{\input{authors_hours_aggregated_expected_29_8.tex}}
\end{minipage}
  \begin{minipage}[c]{0.32\textwidth}
 \resizebox{1\textwidth}{!}{\input{authors_hours_aggregated_expected_22_17.tex}}
\end{minipage}
  \begin{minipage}[c]{0.32\textwidth}
 \resizebox{1\textwidth}{!}{\input{authors_hours_aggregated_expected_13_9.tex}}
\end{minipage}
	\caption{\small{\textbf{Abnormal authors during events - }(Left) Event $(29^{th},7h\text{ - }9h)$: we see that \textit{NicolasSarkozy} is probably responsible for the observed event since its activity deviates significantly from its usual one. (Center) Event $(22^{nd},16h\text{ - }22h)$: the cause of this event is multiple, we see that several authors, mostly individuals and politicians from the right-wing, are more retweeted than they usually are. (Right) Event $(13^{th},9h)$: the distribution is more homogeneous, which makes this event a more global phenomenon.}}
	\label{fig:abnormal_author_hours}
\end{figure*}

\noindent\textit{1) One main author}\\
Figure \ref{fig:abnormal_author_hours} (Left) displays the distribution of deviation values for event $e=(29^{th},7h\text{ - }9h)$. Most observations $d\in \mathcal{D}$ follow a Gaussian distribution centred on $\bar{d_p}=0$. We find 14 abnormal values. Among those, the one corresponding to $(\mbox{NicolasSarkozy},29^{th},7h\text{ - }9h)$ significantly deviates from others. Indeed, in the considered context, we expect NicolasSarkozy to account for $$\frac{v(\cdot,\mbox{NicolasSarkozy},\cdot,\{7h,8h,9h\})}{v(\cdot,\cdot,\cdot,\{7h,8h,9h\})}=2.2\%$$ of all retweets observed from $7h$ to $9h$. Thus, on the $29^{th}$ of August from $7h$ to $9h$, we expect him to be retweeted $v(\cdot,\cdot,(29^{th},7h\text{ - }9h))\times 2.2\%=194$ times. Yet, he was re-tweeted $1,644$ times, which explain its large deviation value. \\
Table \ref{tab:events} lists events with a similar distribution. In most cases, we observe that the corresponding media event is centred on the main author. For instance, they often indicate a political meeting of this author.\\

\noindent\textit{2) Several main authors}\\
Figure \ref{fig:abnormal_author_hours} (Center) displays the distribution of the set of deviation values for event $e=(22^{nd},16h\text{ - }22h)$. Once more, most observations $d\in \mathcal{D}$ follow a Gaussian distribution centred on $\bar{d_p}=0$. We detect $42$ outliers. Several values significantly deviates from the mean, indicating, this time, several main authors.\\ 
These events are not due to a single popular author, but to several authors, considerably retweeted. In contrast to the previous example, this suggest that they originate from the reaction of a few authors to some external fact in which they have an interest. This is what we observe in Table \ref{tab:events}: media events related to events with similar distributions are often indicative of situations according to which the main authors are not related, but on which they react. For example, the event on August the $3^{th}$, on the intervention of the police in a church, and the one on August the $26^{th}$, on burkini wearing, are media events intensely taken up by political members of right and extreme-right wings.\\

\noindent\textit{3) No main authors}\\
Figure \ref{fig:abnormal_author_hours} (Right) displays the distribution of deviation values for event $e=(13^{th},9h)$. In opposition to previous examples, we see that values are more homogeneously distributed and spread over a smaller range.\\
The absence of significant outliers shows that these events are more global phenomena than the previous ones: they emerge because numerous authors are being retweeted instead of a few, intensely. This suggest that they originate from the reaction of a multitude of authors to a general current affair. This is the case, for instance, of the two Olympic victories of France on the $19^{th}$ and $21^{th}$ of August (see Table \ref{tab:events}). \\

Studying interactions by looking at authors enables us to have a deeper understanding of events. In particular, it enables us to identify authors which are unexpectedly and primarily retweeted. This gives us hints on the event's origin: it might results of a focus on a single author, or multiple authors, or none in particular. 


\subsection{Abnormal spreaders during events}
\label{subsec:spreaders}
Among the three previous cases, we are now interested in events generated by a single author (case 1). In particular, we seek to determine if their emergence is due to a large number of spreaders, or on the contrary, if they emerge only because of a small number of spreaders which retweet them abnormally.\\

To do so, we proceed as in the previous section and locally study interactions in the filtered data cube $\mathcal{C}_3(S\times \{a^*\}\times \{e\},v)$, where $a^*$ is the predominant abnormal author corresponding to event $e$. A cell $(s,a^*,e)$ within this cube gives the total number of times $s$ retweeted $a^*$ during $e$. This way, we focus on how each of the spreaders retweeted $a^*$ during the event.\\

Expected values are defined from data cube $\mathcal{C}_4(S\times \{a^*\}\times D\times \{H_e\},v)$, using the comparison data cubes $\mathcal{C}_3(S\times \{a^*\}\times \{H_e\},v)$ and $\mathcal{C}_2(\{a^*\}\times\{H_e\},v)$, obtained by aggregation, and $\mathcal{C}_3(S\times \{a^*\}\times \{e\},v)$ obtained by aggregation and filtering: 
$$ v_{exp}(s,a^*,e)=v(\cdot,a^*,e)\times \frac{v(s,a^*,\cdot,H_e)}{v(\cdot,a^*,\cdot,H_e)},$$
where $v(\cdot,a^*,e)$ is the total number of retweets $a^*$ received during $e$; $v(s,a^*,\cdot,H_e)$ is the total number of time spreader $s$ retweeted author $a$ on hours of $H_e$; and $v(\cdot,a^*,\cdot,H_e)$ is the total number of retweets author $a$ received during $H_e$.\\

According to this context, a triplet $(s,a^*,e)\in S\times \{(a^*,e)\}$ is abnormal because there is a deviation between the number of time $s$ retweeted $a$ during $e$, and the number of time $s$ is expected to retweet $a$ during this same period on other days. Similarly, three situations arise.\\

\begin{figure}
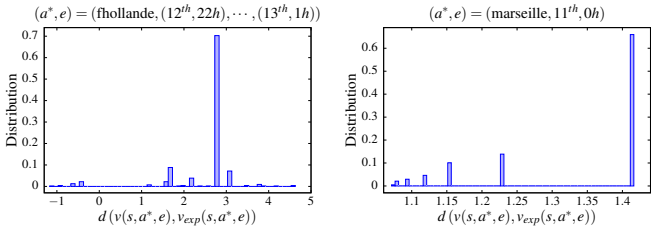

	\begin{subfigure}[t]{0.25\textwidth}
	  \resizebox{0.95\textwidth}{!}{\input{spreaders_fhollande.tex}}
	\end{subfigure}\hspace*{-0.1cm}
	\begin{subfigure}[t]{0.25\textwidth}
	    \resizebox{0.95\textwidth}{!}{\input{spreaders_marseille.tex}}
	    	\end{subfigure}
	    		    	 \caption{\small{\textbf{Distribution of deviation values in the case where all spreaders behave normally --} Bars beyond (\textit{resp.} below) 0 correspond to spreaders which retweet $a^*$ during $e$ more (\textit{resp.} less) than usual. For instance, the most extreme positive value for fhollande corresponds to a spreader which never retweeted fhollande from $22h$ to $1h$, except six times during the event. The most extreme negative value corresponds to a spreader which retweeted him once during the event, even though he retweeted him 7 times in total during this period.}}
	    	\label{fig:fhollande_marseille}
\end{figure}

\noindent\textit{1) Global phenomena}\\
For events $(\text{fhollande}, (12^{th},22h), \cdots, (13^{th},1h))$ and $(\text{mar-}$ $\text{seille}, 11^{th},0h)$, we observe distributions in which the range of deviation values is very small (see Figure \ref{fig:fhollande_marseille}). In the first case, we observe 22 different deviation values. Moreover, $90\%$ of all triplets $(s,a^*,e)$ have their deviation equal to $1.7$, $2.2$, $2.8$, or $3.1$. For marseille, we observe the same patterns: there are only 7 different deviation values, among which $90\%$ of all triplets are distributed between values $1.41$, $1.23$, and $1.16$ (see Figure \ref{fig:fhollande_marseille}). Some of the behaviors corresponding to these values are described in Table \ref{tab:fhollande_marseille}. \\
These distributions show a limited number of spreaders behaviors. None of them have significantly different activities than others. Thus, the emergence of fholland and marseille is due to a global phenomenon in which a large number of spreaders retweeted them.\\

\begin{table}
\hspace*{-0.2cm}
\scalebox{0.85}{
  \setlength\tabcolsep{1.8pt}
\begin{tabular}{|l|c|c|c|c|c|c|c|}
\hline
  \centered{ Event } &
  \multicolumn{4}{c|}{\centered{$((12^{th},22h), \cdots, (13^{th},1h))$}} &
  \multicolumn{3}{c|}{\centered{$(11^{th},0h)$}} \\
  \hline
  \centered{ Abnormal author} &
\multicolumn{4}{c|}{\centered{fhollande}} &
\multicolumn{3}{c|}{\centered{marseille}}  \\
    \hline
  \centered{Deviation value} &
  $\:\:\,\, 1.7\:\:\,\,$ &
  $\:\:\,\, 2.2\:\:\,\,$ &
  $\:\:\,\, 2.8\:\:\,\,$ &
  $3.1$ &
  $\:\:\,\,1.41\:\:\,\,$&
  $\:\:\,\,1.23\:\:\,\,$ &
  $\:\:\,\,1.16\:\:\,\,$ \\
      \hline  
      \centered{ $\%$ of spreaders} &
  \centered{$9$}&
  \centered{$4$}&
  \centered{$70$} &
  \centered{$7$} &
  \centered{$66$}&
  \centered{$14$} &
  \centered{$10$} \\
      \hline
  \centered{ Number of retweets\\during $e$ } &
  \centered{$1$}&
  \centered{$2$}&
  \centered{$1$} &
  \centered{$2$} &
  \centered{$1$}&
  \centered{$2$} &
  \centered{$3$} \\
      \hline
  \centered{ Total Number of retweets\\ from $h_i$ to $h_j$} &
    \centered{$2$}&
  \centered{$3$}&
  \centered{$0$} &
  \centered{$0$} &
  \centered{$0$}&
  \centered{$0$} &
  \centered{$0$} \\
      \hline
\end{tabular}
}
 \caption{\small{\textbf{Most probable behaviors in the case where all spreaders behave normally --} In both cases, the most probable deviation value corresponds to spreaders which retweets $a^*$ only once during the period. For marseille, we observe that the larger the number of retweets during $e$, the smaller the deviation value. This is due to Poisson deviation which takes into account the importance of the deviation between the observed value and its expected one.}}
	    	\label{tab:fhollande_marseille}
\end{table}

\begin{figure*}
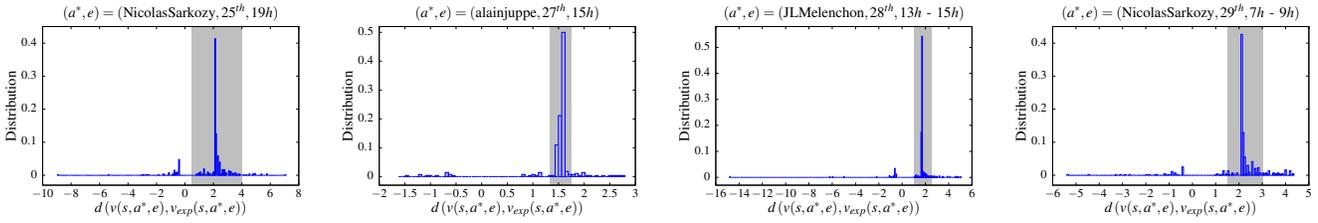

	\begin{subfigure}[t]{0.24\textwidth}
	  \resizebox{0.95\textwidth}{!}{\input{spreaders_NicolasSarkozy_25_19.tex}}
	\end{subfigure}
	\begin{subfigure}[t]{0.24\textwidth}
	    \resizebox{0.95\textwidth}{!}{\input{spreaders_alainjuppe_27_15.tex}}
	    	\end{subfigure}
	    		\begin{subfigure}[t]{0.24\textwidth}
	  \resizebox{0.95\textwidth}{!}{\input{spreaders_JLMelenchon_28_13_14_15.tex}}
	\end{subfigure}
		\begin{subfigure}[t]{0.24\textwidth}
	  \resizebox{0.95\textwidth}{!}{\input{spreaders_NicolasSarkozy_29_7_8_9.tex}}
	\end{subfigure}
\caption{\small{\textbf{A group of spreaders behave abnormally -} In each distribution, similar behaviors are observed. Most spreaders retweet $a^*$ once or twice during $e$ while they usually never retweet $a^*$ at this time of day. These unusual but not significantly deviating behaviors are represented by the Gaussian curve with an average $\bar{d_p}$ between $1.6$ and $2.3$. Those who are used to retweet $a^*$ at this time of day have deviation values either close to 0, if they retweeted as they are used to, negative, if they retweeted less, or positive, if they retweeted more. In this last case, the group of spreaders which behave abnormally is largely responsible for the emergence of $a^*$. }}
	    	\label{fig:groupe_spreaders}
\end{figure*}

\noindent\textit{2) Group of online activists}\\
Figure \ref{fig:groupe_spreaders} shows the distributions of deviation values for ev-ents $(\text{NicolasSarkozy},25^{th},19h)$, $(\text{alainjuppe},27^{th},15h)$,\\ $(\text{JLMelenchon},28^{th},13h\text{ - }15h)$ and $(\text{NicolasSarkozy},29^{th},$\\$7h\text{ - }9h)$. Most observations $d_p\in \mathcal{D}$ follow a Gaussian distribution centred on a mean $\bar{d_p}$. Contrary to distributions in Sections \ref{subsec:hours} and \ref{subsec:authors}, $\bar{d_p}$ varies from $1.6$ to $2.3$. This shift indicates that globally, spreaders have an activity which is higher than usual, which partly explains the emergence of main author $a^*$. \\
We detect negative and positive outliers. Negative outliers indicate spreaders who retweet $a^*$ less that they are supposed to. As such, they do not influence the emergence of $a^*$. On the contrary, positive outliers, who are spreaders more active than usual, play a key role regarding the importance of $a^*$ during $e$. This is what we observe in Table \ref{tab:activist}. For all events, we notice a small group of spreaders which extensively retweets $a^*$ and which accounts for a significant proportion of the total number of retweets. Within this group, several spreaders retweet $a^*$ more than $50$ times during the event. Even if they represent a very small portion of all spreaders, they are a major cause of the emergence of $a^*$ during $e$.   \\

\begin{table*}
  \begin{minipage}[c]{0.57\textwidth}
\setlength\tabcolsep{1.8pt}
\begin{tabular}{|l|c|c|c|c|}
\hline
  \centered{ Event } &
  \centered{$(25^{th},19h)$} &
  \centered{$(27^{th},15h)$} &
  \centered{$(28^{th},13h\text{ - }15h)$} & 
  \centered{$(29^{th},7h\text{ - }9h)$} \\
  \hline
  \centered{ Abnormal\\ Author} &
  \centered{NicolasSarkozy}&
  \centered{alainjuppe} &
  \centered{JLMelenchon} &
  \centered{NicolasSarkozy} \\
    \hline
  \centered{ $\%$ of\\ abnormal\\ spreaders} &
    \centered{$2.7$}&
  \centered{$6.7$} &
  \centered{$4.5$} &
  \centered{$6$} \\
      \hline
        \centered{ $\%$ of\\ retweets } &
            \centered{$14$}&
  \centered{$40$} &
  \centered{$37$} &
  \centered{$25$} \\
      \hline
\end{tabular}
 \end{minipage}
  \begin{minipage}[c]{0.42\textwidth}
\caption{\small{\textbf{Group of influential spreaders -} We observe that a small proportion of spreaders constitutes in fact a significant part of all retweets received by the main author during the event. For instance, for $(\text{alainjuppe},27^{th},15h)$, we detect 19 abnormal spreaders ($6.7\%$ of all spreaders). Together, they retweeted alainjuppe $513$ times at $15h$, which consists in $40\%$ of all its retweets during this hour.}}
\label{tab:activist}
 \end{minipage}
\end{table*}

\noindent\textit{3) One online activist}\\
Event $(\text{FrancoisFillon},11^{th},3h)$ is an extreme case of the previous situation. The group of abnormal spreaders solely consists in one user which retweets FrancoisFillon 73 times at $3h$. Hence, the emergence of FrancoisFillon the $11^{th}$ at $3h$ is only due to this unique spreader which accounts for $100\%$ of all its retweets.\\

Here again, local analysis of spreaders leads us to notice that some events are more global phenomena than others. In particular, some authors emergence is partly due to a small group of spreaders that substantially retweets them, which could mislead other users on the significance of these authors. Thereby, this analysis highlights crucial information that should be taken into account to evaluate the relevance of an event.

\subsection{Abnormal hashtags}
It is possible to gain supplementary information on previous events by adding a content-based dimension using hashtags. In this section, we apply our method on dataset $\mathcal{D}_2$ and focus on the four dimensions: spreaders, authors, hashtags and time. First, we search for hours in which some hashtags are abnormally retweeted, then establish a correlation with previously detected events.\\

We are interested in abnormal triplets $(k^*,d^*,h^*)$ in data cube $\mathcal{C}_3(K \times D \times H,v)$. Given the ephemeral nature of hashtags, we use expected values slightly different than the previous ones. This time, we take into account the expected activity during hour $h$ and we adjust it with the number of hashtags $k$ retweeted on day $d$:
$$ v_{exp}(k,d,h)=v(\cdot,\cdot,k,d,\cdot)\times \frac{v(\cdot,\cdot,\cdot,\cdot,h)}{v(\cdot,\cdot,\cdot,\cdot,\cdot)}.$$
This way, we do not assume that the number of hashtags observed at fixed hours is constant. According to this context, a triplet $(k^*,d^*,h^*)$ is abnormal when there is a significant deviation between the number of retweets containing hashtag $k^*$ during $(d^*,h^*)$, and the number of hashtags $k^*$ that would be retweeted on day $d$ if they were distributed among hours proportionally to their activity.\\

We find 225 abnormal triplets $(k^*,d^*,h^*)$, including 114 different hashtags (by ignoring differences in cases and accents). Among the 225 abnormal triplets, $43\%$ correspond to a previously found abnormal event (in Subsections \ref{subsec:hours}, \ref{subsec:authors} and \ref{subsec:spreaders}). Tables \ref{tab:privi}, \ref{tab:sevprivi}, and \ref{tab:noprivi} display abnormal hashtags according to their corresponding event, for events with respectively one, several and no main author(s). We can make several observations.\\

\begin{table*}
\setlength\tabcolsep{1.8pt}
\begin{tabular}{|l|c|c|c|c|c|}
\hline
  \centered{ Event } &
  \centered{$\left( (12^{th},22h),...,(13^{th},1h)\right) $} & 
  \centered{$(25^{th},19h)$} &
  \centered{$(27^{th},15h)$} &
  \centered{$(28^{th},13h\text{ - }15h)$} &
  \centered{$(29^{th},7h\text{ - }9h)$} \\
    \hline
  \centered{ Abnormal\\ hashtags} &
  \centered{judo\\ rio2016\\ fra\\espritbleu \\\textit{(blue spirit)}} &
  \centered{\textbf{Campaign Slogan:} \\ toutpourlafrance \\ (\textit{all for France})\\ \\ \textbf{Location:}\\ chateaurenard} &
  \centered{\textbf{Campaign Slogan:} \\ 3moispourgagner\\(\textit{3 month to win})} &
  \centered{\textbf{Campaign Slogan:} \\ benoithamon2017\\lagauchepourgagner \\\textit{(left for win)}\\ insoumis28aout \\ (\textit{rebellious of august $28^{th}$})\\\\ \textbf{TV/Radio program:}\\ LeGrandJury } &
  \centered{\textbf{Campaign Slogan:} \\ toutpourlafrance\\\textit{(all for France)} \\\\ \textbf{TV/Radio program:}\\ rtlmatin\\ télématin \textit{(morning show)}\\ bourdindirect\\ invitépol \textit{(political guest)}} \\ 
      \hline
\end{tabular}
\caption{\small{\textbf{Abnormal hashtags of events with one main author.}}}
\label{tab:privi}
\end{table*}

\begin{table*}
\setlength\tabcolsep{2.2pt}
\begin{tabular}{|l|c|c|c|c|c|}
\hline
  \centered{ Event } &
  \centered{$(3^{rd},10h\text{ - }13h)$} &
  \centered{$(22^{th},16h\text{ - }22h)$} &
  \centered{$(24^{th},20h\text{ - }22h)$} &
  \centered{$(26^{th},15h\text{ - }18h)$} &
  \centered{$(28^{th},0h)$} \\
  \hline
  \centered{ Abnormal\\ hashtags} &
   \centered{sainterita \\ \textit{(name of a church)}} &
  \centered{sarkozy \\\\ \textbf{Campaign Slogan:} \\ toutpourlafrance \\\textit{(all for France)}\\\\ \textbf{TV/Radio program:} \\clubdelapresse, e1soir} &
  \centered{sarko \\\\ \textbf{Campaign Slogan:} \\ toutpourlafrance \\\textit{(all for France)} \\\\ \textbf{TV program:} \\ ns20h } &
  \centered{burkini\\ conseildetat \\ \textit{(council of state)}\\\\ \textbf{TV/Radio program:} \\ BFMTV} &
  \centered{salafisme \textit{(salafism)}\\\\ \textbf{TV program:} \\ ONPC } \\
      \hline
\end{tabular}
\caption{\small{\textbf{Abnormal hashtags of events with several main authors.}}}
\label{tab:sevprivi}
\end{table*}

\begin{table*}
  \begin{minipage}[c]{0.68\textwidth}
\setlength\tabcolsep{1.8pt}
\begin{tabular}{|l|c|c|c|c|c|}
\hline
  \centered{ Event } &
  \centered{$(13^{th},9h)$} &
  \centered{$(19^{th},22h)$} & 
  \centered{$(21^{st},21h)$} &
  \centered{$(23^{th},7h\text{-}8h)$} & 
  \centered{$(30^{th},17h\text{-}18h)$} \\
  \hline
  \centered{ Abnormal hashtags} &
  \centered{$\diagup$}&
    \centered{rio2016} &
  \centered{rio2016\\ boxe \textit{(boxing)}} &
  \centered{$\diagup$} &
  \centered{macron} \\
      \hline
\end{tabular}
 \end{minipage}
  \begin{minipage}[c]{0.31\textwidth}
\caption{\small{\textbf{Abnormal hashtags of events with no main authors.}}}
\label{tab:noprivi}
 \end{minipage}
\end{table*}

First, we notice that an event is often attached to a political slogan together with a radio or television show. In this case, there are three possible situations: either the show receives a political guest, or the show speaks about a topicality associated with one or more politician(s), or on the opposite, the show and the political slogan are uncorrelated -- for instance, in the case in which several current events happen within the same period.\\

We notice that events in Tables \ref{tab:sevprivi} and \ref{tab:noprivi} are always associated with a general term, independent from a political slogan or a show. As suggested by the analysis of anomalous authors, this shows that the corresponding event results from the reaction to an external fact. For instance, hashtags "\textit{Rio2016}" are related to the global reaction of users to Olympic victories of France. Hashtag "SainteRita", on the other hand, is related to the reaction of users to a police intervention in a church. Furthermore, events $(22^{nd},16h\text{ - }22h)$ and $(24^{th},20h\text{ - }22h)$, attached to hashtags "\textit{Sarkozy}" and "\textit{Sarko}", suggest that there is a discussion about Nicolas Sarkozy apart from official tweets and hashtags released by his team. In particular, on the $22^{nd}$, people react to the announcement of Nicolas Sarkozy's candidacy to presidency: this event corresponds with the first use of hashtag "\textit{ToutpourLaFrance}" which is his campaign slogan.\\


We observe another interesting fact: on the $28^{th}$ from $13h$ to $15h$, we detect the campaign slogan of JLMelenchon, "\textit{insoumis28aout}", which is expected since JLMelenchon is the predominant author of this event. However we also detect campaign slogans of benoithamon, another politician -- "benoithamon2017" and "LaGauchePourGagner" -- which is unexpected since it does not appear as a predominant author in the previous study. \\

Finally, we notice that events $(11^{th},0h)$, $(11^{th},3h)$, $(13^{th},$\\ $9h)$ and $(23^{th},7h\text{-}8h)$ are not related to any detected hashtags. This is due to the fact the analysis performed in this subsection is global. With local analysis of abnormal hashtags, centred on events, as done before with authors in Subsection \ref{subsec:authors}, we succeed in identifying the corresponding discussed topics. For instance, during event $(13^{th}, 9h)$, we identify abnormal hashtags \textit{etatdurgence} (\textit{state of emergency}), \textit{cazeneuve} and \textit{islamigration}, referring to a measure taken that same day by the minister of the Interior, Bernard Caze-neuve. \\ 

In this section, we applied our method to datasets $\mathcal{D}_1$ and $\mathcal{D}_2$. We detected abnormal events, independent of the activity of the day or time considered. Then, we performed local analysis on each of these events, using numerous different contexts, more or less filtered or aggregated. This allowed us to understand their emergence. For instance, we learned that on the $11^{th}$ at $3h$, one unique spreader intensely retweets FrancoisFillon; that from the $12^{th}$ at $22h$ to the $13^{th}$ at $1h$, numerous spreaders retweet fhollande once, regarding an Olympic victory of France in judo; or, that on the $27^{th}$ at $15h$, a small group of spreaders is largely responsible for the emergence of alainjuppe during its political meeting. Our method provides the possibility of studying further aspects of interactions by choosing new relevant contexts. In the following section, we discuss two other possible applications.

\section{Other Applications}
\label{sec:discussion}

Observations made in the previous section open up several research perspectives. On the one hand, given the ubiquity of news related hashtags within each events -- as TV and Radio programs --, it would be interesting to characterize more precisely the reaction of users to television shows through Twitter. On the other hand, we could focus on topic dynamic over time and, in particular, on prediction of user-topic links.

\subsection{Characterization of second screen usage}
\label{sec:second_screen}
The characterization of second screen usage is a very recent field of study. The term \textit{second screen} refers to a web-connected screen, like a smartphone or a laptop, that people use to comment about TV programs on social media while watching television. As part of this study, it is interesting to analyse the differences between what is said in the TV program and the ensuing discussions on Twitter. This has been applied in many situations, in particular, to follow sport events \cite{corney2014spot} and political debates \cite{giglietto2014second,freelon2015big,gil2015second}. In the following, we provide elements to characterize the second screen usage with our method. This is a novel approach since previous studies often consist either in manual comparison between tweets content and a record of the discussion that took place in the TV show, or in a focus on the television audience or the number of tweets over time. We focus on Nicolas Sarkozy's appearance on television news for the launch of his campaign, on the $24^{th}$ of August from $20h$ to $22h$. \\

\begin{figure}
 \resizebox{0.5\textwidth}{!}{\input{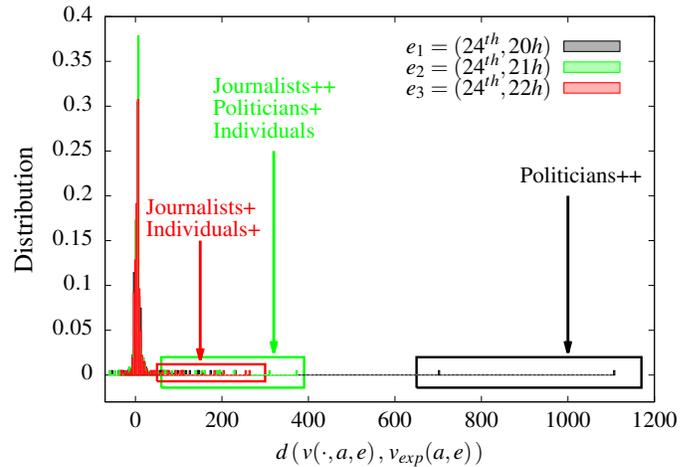}}
	\caption{\small{\textbf{Evolution of abnormal authors distributions on the $24^{th}$ of August from $20h$ to $22h$. }}}
	\label{fig:second}
\end{figure}

First, we focus on abnormal authors using the same expected values as in Section \ref{subsec:authors}, separately on each hour. Figure \ref{fig:second} displays the distribution of the set of deviation values for $e_1=(24^{th},20h)$, $e_2=(24^{th},21h)$ and $e_3=(24^{th},22h)$. At $20h$, there are two predominant authors: \textit{NicolasSarkozy} and \textit{TTpourlaFrance}, his team's account. At $21h$, another situation occurs. The set of values is more homogeneous: there are more outliers, but less significant. Among these, we see many journalists as well as right-wing politicians supporting Nicolas Sarkozy. We notice that some people, neither related to a newspaper nor to a political team, begin to appear among abnormal authors. Finally, at $22h$, the range of values is even smaller, meaning that the observed event is not the result of a focus on a limited number of authors, but a global phenomenon where everyone retweets everyone. Among outliers, we only see journalists and anonymous users. Hence, the more time passes, the more distributions are homogeneous, showing that the event becomes a global phenomenon as information spread.\\

The previous analysis shows that the political interview on television is taken up by users on social media. In the same way as with abnormal authors, we now focus on abnormal hashtags to analyse how the discussion evolves over time. We observe similar distributions. At $20h$, the two hashtags \textit{ns20h} and \textit{toutpourlafrance} point out strongly. At $21h$ and $22h$, distributions are more homogeneous. The two previous hashtags released by Nicolas Sarkozy's team are still abnormal at $21h$, but become normal again at $22h$. Other hashtags are abnormal only from $21h$ to $22h$ or from $22h$ to $23h$. Among these, we find terms used by Nicolas Sarkozy during his interview, such as \textit{chomage} (\textit{unemployment}). Finally, we observe an evolution of hashtags referring to the same topic: at $21h$, \textit{hollande}, then at $22h$, \textit{hollandedemission} (\textit{hollande resignation}); or \textit{schengen} at $21h$, then \textit{stopschengen} at $22h$; or \textit{burkini} from $20h$ to $22h$, then \textit{bikini} from $22h$ onwards. \\

This preliminary analysis could be continued. For instance, when studying abnormal hashtags, we could use local contexts, restrained to journalists, or Nicolas Sarkozy's political team, or independent users, in order to analyse which hashtags each of these communities propagate. Also, we could focus on the evolution of hashtags belonging to a same topic and see if they are retweeted by the same community of spreaders.

\subsection{Predicting User-Topic Links}
The latter question attract a lot of interest among researchers: many are interested in topic dynamics and in particular, predicting user-topic links. The first difficulty lies in finding the set of terms forming a topic, \textit{i.e.} a consistent semantic content. Some researchers characterize it from a set of hashtags whose temporal evolutions are similar \cite{pepin2015visual}, or from clusters of hashtags which are highly associated within tweets \cite{cardoso2017topical}. Others use text processing techniques to infer a topic from the entire text within tweets, rather than only using hashtags \cite{xie2016topicsketch}. To predict user-topic links, most researchers use machine learning techniques for sentiment analysis \cite{ren2013predicting,si2013exploiting,colleoni2014echo,reyes2018understanding}. We also find methods based on lexicon \cite{o2010tweets}. In the following, we propose a new approach which consist in finding topics among abnormally retweeted hashtags. \\

We only have the structure of retweets $(s,a,k,d,h)$. In order to identify topics from this data, we take advantage of the fact that users are engaged in a cause, especially in the case of political communication. That is, an author will often post tweets related to this cause, and spreaders committed to the cause will retweet them intensely. Thus, we define a topic as being a set of hashtags retweeted intensely by the same spreaders and for which a common group of authors is intensely retweeted. \\

\begin{figure}
\hspace*{0.4cm}
\begin{tikzpicture}

\fill [green!75!black] (1.25,2.5) rectangle (3.25,3.25);
\fill [green!75!black] (4.75,2.5) rectangle (6.75,3.25);

\fill [orange!40!white] (0.7,1.25) rectangle (2.7,2);
\fill [orange!70!white] (2.95,1.25) rectangle (4.95,2);
\fill [orange!70!white] (5.2,1.25) rectangle (7.3,2);

\fill [blue!40!white] (0,0) rectangle (2,0.75);
\fill [blue!70!white] (3,0) rectangle (5,0.75);
\fill [blue!70!white] (6,0) rectangle (8,0.75);

\node[draw, fit={(1.25,2.5) (3.25,3.25)}, inner sep=0pt, label=center:$a^*_1$] (A) {};
\node[draw, fit={(4.75,2.5) (6.75,3.25)}, inner sep=0pt, label=center:$a^*_2$] (B) {};

\node[draw, fit={(0.7,1.25) (2.7,2)}, inner sep=0pt, label=center:$k_1$] (C) {};
\node[draw, fit={(2.95,1.25) (4.95,2)}, inner sep=0pt, label=center:$k_2$] (D) {};
\node[draw, fit={(5.2,1.25) (7.3,2)}, inner sep=0pt, label=center:$k_3$] (E) {};

\node[draw, fit={(0,0) (2,0.75)}, inner sep=0pt, label=center:$s^*_1$] (F) {};
\node[draw, fit={(3,0) (5,0.75)}, inner sep=0pt, label=center:$s^*_2$] (G) {};
\node[draw, fit={(6,0) (8,0.75)}, inner sep=0pt, label=center:$s^*_3$] (H) {};

\draw (A)--(C);
\draw (A)--(D);
\draw (A)--(E);

\draw (B)--(C);
\draw (B)--(D);
\draw (B)--(E);

\draw (F)--(C);
\draw (G)--(D);
\draw (G)--(E);
\draw (H)--(E);

\end{tikzpicture}
\caption{\small{\textbf{Formation of topics from hashtags --} For $K_3=\{k_1,k_2,k_2\}$, $S^*_{k_1}=\{s^*_1\}$, $S^*_{k_2}=\{s^*_2\}$, and $S^*_{k_3}=\{s^*_2, s^*_3\}$. Then, $S^*_{K_3}=\emptyset$ and $K_3$ does not constitute a topic. On the other hand, $K_2=\{k_2,k_2\}$ is a topic since $A^*_{K_2}=\{a^*_1,a^*_2,a^*_3\}$ and $S^*_{K_2}=\{s^*_2\}$.}}
\label{fig:topics}
\end{figure}
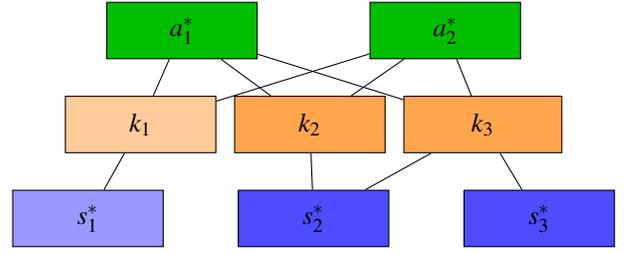

Formally, let $K_N \subseteq K$ be a set of $N$ hashtags. We proceed as follows. First, for each hashtag $k_i\in K_N$, we locally search what are the abnormal spreaders associated to $k_i$ according to the following expected values
$$v_{exp}(s,k_i,d,h)=v(\cdot,\cdot,k_i,d,h) \times \frac{v(s,\cdot,\cdot,\cdot,h)}{v(\cdot,\cdot,\cdot,\cdot,h)}\quad.$$
We obtain an abnormal spreader group denoted $S^*_{k_i}$ such that $s\in S^*_{k_i}$ is a spreader that retweets hashtag $k_i$ abnormally during a specific hour, given its usual activity at this time of the day. After performing this step on all hashtags, we define the group of spreaders related to $K_N$ as the set of abnormal spreaders common to all hashtags in the set: $S^*_{K_N}=\bigcap_{i=1}^{i=N} S^*_{k_i}$. We proceed symmetrically to find the set of abnormal authors related to $K_N$, denoted $A^*_{K_N}$. Given the set of abnormal spreaders and authors related to $K_N$, we say that $K_N$ is a topic if both $S^*_{K_N}$ and $A^*_{K_N}$ are non-empty (see Figure \ref{fig:topics} for illustration). Note that we are only interested in abnormal authors and spreaders since they are the ones which unquestionably want to propagate the topic.\\

\begin{figure*}
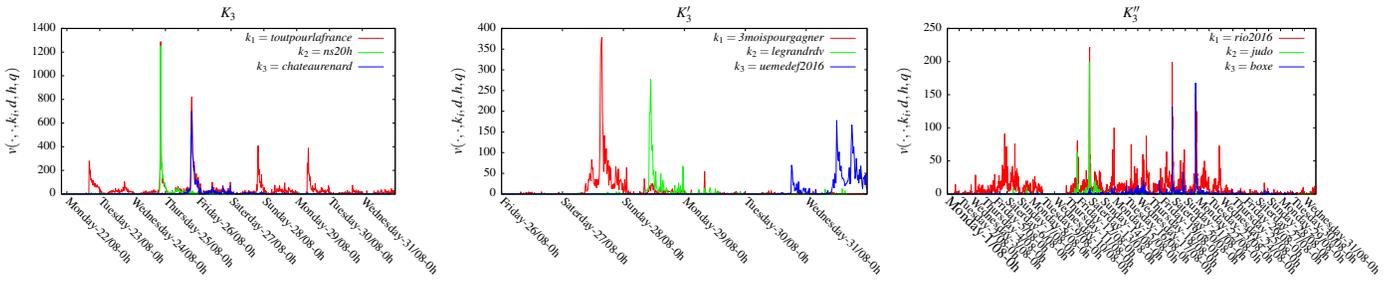

  \begin{minipage}[c]{0.32\textwidth}
 \resizebox{1\textwidth}{!}{\input{evolution_k3.tex}}
\end{minipage}
  \begin{minipage}[c]{0.32\textwidth}
 \resizebox{1\textwidth}{!}{\input{evolution_k2.tex}}
\end{minipage}
  \begin{minipage}[c]{0.32\textwidth}
 \resizebox{1\textwidth}{!}{\input{evolution_k1.tex}}
\end{minipage}
	\caption{\small{\textbf{Evolution of the number of retweets containing hashtag $k_i$ for three different topics -- } Notice that, in order to have a better accuracy, we plotted the number of retweets containing hashtag $k_i$ per quarter $q$. We see that hashtags dynamics within a same topic can be uncorrelated as in $K^\prime_3$, or correlated as \textit{ns20h} and \textit{chateaurenard} with \textit{toutpoutlafrance} in $K_3$.}}
	\label{fig:topic_evol}
\end{figure*}

With $N = 3$ and by considering the set of triplets obtained from the 114 abnormal hashtags identified in the previous section, we find $876$ topics. For instance, we identify topic 
$K_3=\{\text{\textit{chateaurenard, ns20h,toutpourlafrance}}\}$, which has 4 abnormal authors belonging to the same political party, $A^*_{K_3}=\{\text{{\small GilAverous, LArribage, NicolasSarkozy, TTpourlaFrance}}\}$,\\ and 48 abnormal spreaders; topic $K^{\prime}_3=\{\text{\textit{3moispourgagner,}}$ $\text{\textit{legrandrdv (radio program), uemedef2016}}$ $\text{\textit{(summer school}}$\\ $\text{\textit{of the employers' federation of France)}}\}$ associated to one abnormal author, alainjuppe, and to a group of $18$ abnormal spreaders; and topic $K^{\prime\prime}_3=\{\text{\textit{boxe (boxing), judo, rio2016}}\}$ associated to $7$ abnormal authors from different origins, and only $3$ abnormal spreaders\footnote{Note that in this case, we only find 3 abnormal spreaders since events related to sport are usually homogeneous events which do not exhibit groups of active spreaders.}. Figure \ref{fig:topic_evol} shows the temporal evolution of each hashtag in each topic. We see that hashtags belonging to the same topic do not necessarily have the same dynamics. \\

After this step and from this set of topics, we can infer user's communities according to which topic they are used to retweet or being retweeted. Now, we address the problem of predicting user-topic links. More precisely, we want to predict the number of interactions between spreader $s$, in community $c_s$, and topic $K_N$ during hour $(d,h)$. Link prediction is inextricably related to abnormal link detection. Indeed, if the detection of abnormal quadruplets $(s, K_N, d, h)$ is based on measuring the deviation between an observed value $v(s, K_N, d, h)$ and its expected value $v_{exp}(s , K_N, d, h)$, link prediction focuses on describing normal behavior and therefore, is based on expected values only. For instance, we could predict the number of interactions between $s$ and $K_N$ during $(d,h)$ as
$$v_{exp}(s , K_N, d, h)=$$
$$\frac{v(c_s,\cdot,K_N,\cdot,\cdot)}{v(\cdot,\cdot,K_N,\cdot,\cdot)} \times  \frac{v(s,\cdot,\cdot,\cdot,h)}{v(c_s,\cdot,\cdot,\cdot,h)} \times \frac{v(\cdot,\cdot,K_N,d,h)}{|D|}$$
$$\quad \quad \quad(1) \quad \quad \quad \quad \quad  \: (2)  \quad \quad \quad \quad\quad\:(3)\quad\quad \quad$$
which takes into account $(1)$ the activity of $s$'s community towards topic $K_N$, $(2)$ the activity of $s$ within its community during the hour of the day $h$, and $(3)$ the expected number of retweets of $K_N$ during hour $h$ of day $d$. \\

This prediction can be improved by taking into account the behavior of authors that $c_s$ is used to retweet, towards topic $K_N$. Also, if $K_N$ is a new topic, we could imagine to replace the activity of topic $K_N$ by the mean activity of a set of related topics.\\

Thus, our method may be useful in many empirical studies and applications. In turn, these applications provide feedback and questions necessary to create more and more complex and relevant contexts and thus, take advantage of the scope of possibilities offered by our method.  



\section{Conclusion}
\label{sec:Conclusion}
In this paper, we provided a method to meticulously explore millions of interactions and find unexpected behaviors under a multitude of situations. We applied it in the context of politics, where the stakes to unravel relevant information in the flow of data are particularly high. We showed that our method successfully highlights events and provide explanations for their emergence. In particular, we found abnormally retweeted authors, groups of very active spreaders, and hot topics during the corresponding abnormal periods. Hence, our method highlights crucial information that should be taken into account to evaluate an event reliability on Twitter.\\
One interesting perspective that could be considered would be to aggregate the base cuboid over authors, spreaders or hashtag (or topics) partitions. This would allow us to study each community separately -- especially the ones corresponding to political parties; the relationship they have with each other; as well as the one they have with the different hastags (\textit{resp.} topics). This in turn would enable us to gain insights about communication strategies deployed by each political parties. \\
Moreover, our method applies to temporal networks modelling entities interacting over time in general. Hence, as discussed in Section \ref{sec:discussion}, numerous applications can benefit from it, as for instance, the characterization of second screen usage on social media (\textit{e.g.} Facebook or Twitter) and link prediction (\textit{e.g.} in IP traffic or e-mail exchanges). \\

\section*{Acknowledgements}
This work is funded in part by the European Commission H2020 FETPROACT 2016-2017 program under grant 732942 (ODYCCEUS) and by the ANR (French National Agency of Research) under grants ANR-15- E38-0001 (AlgoDiv).

\bibliographystyle{abbrv}
\bibliography{biblio_article_ip.bib}

\end{document}